\pdfoutput=1
\documentclass[sigconf, 
]{acmart}
\setcopyright{rightsretained}
\usepackage[utf8]{inputenc}
\usepackage[T1]{fontenc}
\usepackage[scaled=0.8]{beramono}
\usepackage{listings}
\usepackage{caption}
\usepackage{subfig}
\usepackage{xcolor}
\usepackage{hyperref}
\usepackage{tikz}
\usepackage{lipsum,adjustbox}
\usepackage{soul}
\usepackage{graphicx}
\usepackage{amsmath}
\usepackage{xspace}
\usepackage{todonotes}
\usepackage{wrapfig}
\usepackage{algorithm2e}
\usepackage[capitalise]{cleveref}
\usepackage{csquotes}
\usepackage{siunitx}
\newcommand{\Legion}{\textsc{Legion}\xspace}

\newcommand{\TraceJump}{\textsc{TraceJump}\xspace}
\newcommand{\QuickSampler}{\textsc{QuickSampler}\xspace}
\newcommand{\BenchExec}{\texttt{BenchExec}\xspace}
\newcommand{\APPF}{\textsc{APPFuzzing}\xspace}
\newcommand{\Angr}{\texttt{angr}\xspace}
\newcommand{\Claripy}{\texttt{claripy}\xspace}
\newcommand{\Python}{\texttt{Python\,3}\xspace}
\newcommand{\union}{\cup}
\newcommand{\bparagraph}[1]{\textbf{#1}} %{\noindent\textbf{#1}}
\newcommand{\prebparagraph}{} %{\\[-0.8em]}

\newcommand{\FIXME}[1]{\colorbox{yellow}{\bf FIXME: }{\bf #1}}

\lstdefinestyle{customc}{
  breaklines=true,
  frame=none,
  language=C,
  numbers=left,
  showstringspaces=false,
  basicstyle=\scriptsize\ttfamily,
  keywordstyle=\bfseries\color{blue},
  commentstyle=\slshape\color{red!50!black},
  identifierstyle=\color{black},
  stringstyle=\color{orange},
}

\usetikzlibrary{calc,positioning, shapes.geometric, fit, backgrounds, patterns, decorations.pathreplacing}

\usetikzlibrary{calc,positioning, shapes.geometric}
\graphicspath{{Figures/}}
\makeatletter
\def\input@path{{Figures/}{Sections/}}
\makeatother
\newif\ifnew
\newtrue
\begin{document}

\title{{\textsc{Legion}: Best-First Concolic Testing}}

\author{Dongge Liu}
\authornote{This research was supported by Data61 under the Defence Science and Technology Group's Next Generation Technologies Program.}
\affiliation{
  \institution{The University of Melbourne}  
  \department{School of Computing and Information Systems}
  \city{Melbourne}
  \state{Victoria}
  \postcode{3053}
  \country{Australia}
}
\email{donggel@student.unimelb.edu.au}

\author{Gidon Ernst}
\affiliation{
  \institution{LMU Munich}  
  \department{Software and Computational Systems Lab}
  \city{Munich}
  \state{Bavaria}
  \postcode{80539}
  \country{Germany}
}
\email{gidon.ernst@lmu.de}

\author{Toby Murray}
\affiliation{
  \institution{The University of Melbourne}  
  \department{School of Computing and Information Systems}
  \city{Melbourne}
  \state{Victoria}
  \postcode{3053}
  \country{Australia}
}
\email{toby.murray@unimelb.edu.au}

\author{Benjamin I.P. Rubinstein}
\affiliation{
  \institution{The University of Melbourne}  
  \department{School of Computing and Information Systems}
  \city{Melbourne}
  \state{Victoria}
  \postcode{3053}
  \country{Australia}
}
\email{benjamin.rubinstein@unimelb.edu.au}

\begin{abstract}
  % \Legion is a grey-box coverage-based concolic tool that aims to balance 
  % the complementary nature of fuzzing and symbolic execution to achieve the best of both worlds.
  % It proposes a variation of \textit{Monte Carlo tree search (MCTS)}
  % that formulates program exploration as sequential decision-making under uncertainty guided by the best-first search strategy.
  % It relies on \emph{approximate path-preserving fuzzing}, 
  % a novel instance of constrained random testing,
  % which quickly generates many diverse inputs that likely target program parts of interest.
  % In Test-Comp~2020~\cite{TESTCOMP20}, the prototype performed within~90\% of the best score in~9 of~22 categories.

  % {\color{orange} Legion combines concolic execution and fuzzing in a principled way}
  
  % {\color{orange} Legion leverages MCTS}
  
  % {\color{orange} Legion proposes APPF}
  
  % {\color{orange} Legion's performance.}

  Concolic execution and fuzzing are two complementary
  coverage-based testing techniques.  How to achieve the best of both remains an open challenge. %One reason is lacking a universal algorithm that can balance their trade-offs in program exploration under different code structures. 
  To address this research problem, we propose and evaluate \Legion.
  \Legion re-engineers the \textit{Monte Carlo tree search (MCTS)} framework from the AI literature to treat automated test generation as a problem of sequential
  decision-making  under uncertainty. Its best-first search strategy provides
  a principled way to learn the most promising program states to investigate
  at each search iteration, based on observed rewards from previous iterations.
  \Legion incorporates a form of
  directed fuzzing that we call \emph{approximate path-preserving fuzzing} (\APPF) to
  investigate program states selected by MCTS. \APPF serves as the Monte
  Carlo simulation technique and is implemented by extending prior
  work on constrained sampling.  We evaluate \Legion against competitors on 2531 benchmarks from the coverage category of Test-Comp~2020, as well as measuring its sensitivity to hyperparameters, demonstrating its effectiveness on a wide variety of input programs.

\end{abstract}


\copyrightyear{2020}
\acmYear{2020}
% \setcopyright{rightsretained}
\acmConference[ASE '20]{35th IEEE/ACM International Conference on Automated Software Engineering}{September 21--25, 2020}{Virtual Event, Australia}
\acmBooktitle{35th IEEE/ACM International Conference on Automated Software Engineering (ASE '20), September 21--25, 2020, Virtual Event, Australia}
\acmDOI{10.1145/3324884.3416629}
\acmISBN{978-1-4503-6768-4/20/09}

\maketitle

\section{Introduction}\label{sec: Introduction}

\begin{displayquote}
  \textit{Theseus killed Minotauros in the furthest section of the labyrinth and then made his way out again by pulling himself along the thread.} ---Pseudo-Apollodorus, Bibliotheca E1. 7 - 1. 9 trans. Aldrich
\end{displayquote}

The complexity of modern software programs are like labyrinths for software testers to wander:
their program states and execution paths form a confusing set of connecting rooms and paths. 
Like the Minotaur, faults often hide deep inside. One might guess at a fault's
possible location via static analysis,
but in order to slay it Theseus needs to know the path to it for sure, and
the software tester needs to know which input will trigger it.

In the myth of Theseus, the hero king finds Minotauros by accurately tracing past paths with
a ball of thread, allowing him to learn and estimate the maze structure.
We argue that the very same tricks, namely recording exact concrete execution
traces 
and applying machine learning to estimate software structure and guide its
exploration, 
can also benefit coverage-based testing.

The focus of this paper is the quest of coverage-based testing, which
is to cover as many paths in
as little time as possible,
delegating Minotaur detection to separate tools (e.g. AddressSanitiser \cite{AddressSanitizer}, UBSan~\cite{ryabinin2014ubsan}, Valgrind~\cite{nethercote2007valgrind}, Purify~\cite{reed1991purify}).
Traditional methods for coverage-based testing have been dominated by
the two complimentary approaches of concolic execution (as exemplified by DART~\cite{DART} and SAGE~\cite{SAGE}) and coverage-guided greybox fuzzing (as exemplified by libFuzzer~\cite{serebryany2015libfuzzer}, AFL~\cite{AFL}, its various extensions such as AFLFast~\cite{bohme2017AFLFast}, AFLGo~\cite{bohme2017AFLGo}, CollAFL~\cite{gan2018collafl}, Angora~\cite{chen2018angora}.

Continuing the mythological metaphor, 
with concolic execution one spends a long time rigorously planning each path through the maze via constraint solving, to make the correct turn at each branching
point and ensure that no path will ever be repeated. 
However, such computation is expensive and, 
for most modern software, the maze is so large that repeating it for
\emph{every} path is infeasible.

In contrast, a coverage-guided fuzzer like AFL blindly scurries around
the maze, neither spending much time on planning nor accurately memorising
the paths and structure traversed. Thus much time is inevitably spent
unnecessarily repeating obvious execution paths.

Observing the complementary nature of these two methods, 
our research aims to generalise them with Theseus's strategy. Our tool
\Legion\footnote{The name is a homage to the Marvel fictional character who changes personalities for different needs. Our strategy can adjust its exploration preference under different metrics.} traces observed execution paths to estimate the maze structure.
Then it identifies the most promising location to explore next, plans its path to reach that location, 
and applies a form of directed fuzzing to explore extensions of the path to that
location.
\Legion precisely traces 
each concrete execution (i.e.\ fuzzing run)
and gathers statistics to refine its knowledge of the maze to inform its decisions
about where to explore next.

Statistics gathering is central to \Legion's approach, unlike traditional
fuzzers like AFL that eschew gathering detailed statistics to save time.
Instead, \Legion aims to harness the power of modern machine learning
algorithms, informed by detailed execution traces as in other contemporary
tools~\cite{VeriFuzz}.

Real-life coverage testing is more complicated than the Theseus myth, as it
requires a universal strategy
that can adjust itself according to different program structures:
i.e.\ to fuzz the program parts that are more suitable to fuzz, and favour
concolic execution elsewhere.
However, how to determine the best balance between these two strategies
remains an open question.

To address this challenge, \Legion adopts the
\textit{Monte Carlo tree search}~(MCTS) algorithm from machine learning~\cite{kocsis2006bandit}.
MCTS  has proven to work well in complex games like Go~\cite{silver2016alphago}, multiplayer board games~\cite{Catan, MagicTheGathering} and poker~\cite{van2009monte}, as it can adapt to
the game it is playing via iterations of simulations and reward analysis.
Specifically,
MCTS learns a policy by successively simulating different plays
and tracking the rewards obtained during each. In \Legion, plays correspond to
concrete execution (directed fuzzing), while rewards correspond to increased
coverage (discovering new execution paths). 
More importantly, MCTS's guiding principle of \textit{optimism in the face of uncertainty} is appropriate for exploring a maze with an unknown structure, randomness, and large branching factors where rigorously analysing
every detail is infeasible.
Instead, MCTS balances \emph{exploitation} of the branches that appear to be most rewarding based on past experience, against \emph{exploration} of less-well
understood parts of the maze where rewards (path discovery) are
less certain.

With the two tricks of Theseus, \Legion provides a principled framework to generalise and harness the complementary
strengths of concolic execution and fuzzing. In doing so, it sheds new light
on the key factors for efficient and effective input search strategies for
coverage-based testing, on diverse software structures.

We term \Legion's directed fuzzing approach \emph{approximate path-preserving
  fuzzing} (\APPF), and are inspired by \QuickSampler~\cite{QuickSampler}.
\APPF aims to fuzz a specific
location of the search space by launching several binary executions that,
with high probability, follow the same path to that location (but might take
different paths afterwards). It is described in~\cref{sec: APPFuzzing algorithm}.

\Legion treats coverage-based testing as progressively exploring a large space with uncertainty, by iteratively sampling inputs using
cheap but less accurate \APPF that targets specific
program states selected by MCTS. Symbolic state information (i.e.\
a path constraint) is required to
seed \APPF for each program state. However, 
\Legion avoids unnecessary
symbolic execution (and constraint solving) by performing symbolic
execution \emph{lazily}: specifically, \Legion computes symbolic successor
states only for program states deemed promising (i.e.\ those for which
existing statistics indicate are worthy of investigation). 

\Legion's adoption of MCTS is designed to ensure that computational power is used to
explore the most beneficial program locations, as determined by
the score function.
In \Legion scores are evaluated according to a modularised reward heuristic,
with interchangeable coverage metrics as discussed in~\cref{sec: a reward evaluation heuristic}.

Our contributions are:
\begin{itemize}
  % \item It accentuates why the key characteristics of MCTS is a perfect fit to concolic testing (Section \ref{sec: The key idea}).
  \item We propose a variation of Monte Carlo tree search (MCTS) that maintains a balance between concolic execution and fuzzing, to mitigate the path explosion problem of the former and the redundancy of the latter (\cref{sec: Legion MCTS}).
  \item We propose \emph{approximate path-preserving fuzzing}, which
    extends a constrained sampling technique, \QuickSampler~\cite{QuickSampler}, to generate inputs efficiently
    that preserve a given path with high probability (\cref{sec: APPFuzzing algorithm}).
  % \item To show its versatility, we instantiated our strategy with one reward evaluation heuristic \ref{sec: a reward evaluation heuristic}.
  \item We conduct experiments to demonstrate that \Legion is competitive
    against other state-of-the-art approaches (\cref{sec: peer competition results}), and evaluate the effect of different MCTS hyperparameter settings.
  % \item Although this paper separates error detection from its scope {\color{blue} like other coverage-based techniques}, Sec. \ref{sec: Future Work} shows how \textsc{Legion} allows using the information from error detectors.
  % \item {\color{red} Under the same multi-armed bandit {\color{blue} strategy}, it can be easily modified to modularise path preserving fuzzing techniques as arms (Section \ref{sec: Future Work}). This allows the strategy to choose the most efficient fuzzer at different program locations based on their performance.}
  % \item {\color{red}  It can be used to compare and evaluate different implementations of symbolic execution and approximate path preserving fuzzers.}
\end{itemize}

\section{Overview} \label{sec: Overview}

% \subsection{Two complementary approaches} \label{sec: two complementary approaches}

\Legion generalises the two traditional approaches to coverage-based testing:
concolic execution and coverage-guided fuzzing.

Concolic execution relies on a constraint solver to generate concrete inputs that can traverse a particular path of interest. New paths are selected by
flipping path constraints of previously-observed execution traces, thereby
attempting to cover all feasible execution paths.
However, it suffers from the high computation cost of constraint solving and
exponential path growth in large applications.

Coverage-guided fuzzing has become increasingly popular during the past decade due to its simplicity and efficiency \cite{DeepHunter, bohme2017AFLFast}.
It generates randomised concrete inputs at low cost (e.g.\ via bit flipping),
by mutating inputs previously observed to lead to new execution paths.
However, the resulting inputs more often than not fail to uncover new
execution paths, nor to satisfy complex constraints to reach deep program
states.

The complementary nature of these two techniques~\cite{Driller},
which \Legion's design harnesses, is highlighted by considering the exploration of
the program \texttt{Ackermann02} in \cref{lst:ackermann02}. This program
is drawn from the Test-Comp~2020 benchmarks\footnote{https://github.com/sosy-lab/sv-benchmarks/tree/testcomp20}~\cite{TESTCOMP20}.

\begin{figure*}[ht]
  \qquad \qquad \begin{minipage}[b]{0.35\textwidth}
\begin{lstlisting}[style=customc]
int ackermann(int m, int n) {
  if (m==0) return n+1;
  if (n==0) return ackermann(m-1,1);
  return ackermann(m-1,ackermann(m,n-1));
}

void main() {
  int m = input(), n = input();
  // choke point
  if (m < 0 || m > 3) || (n < 0 || n > 23) {
    log(n,m);               // common branch
    return;
  } else {
    int r = ackermann(m,n); // rare branch
    assert(m < 2 || r >= 4);
  }
}

\end{lstlisting}
  % \vspace*{-.3cm}
  \caption{\texttt{Ackermann02.c}}
  \label{lst:ackermann02}
  \end{minipage}
  \hfill
  \begin{minipage}[b]{0.5\textwidth}
  \begin {tikzpicture}[
  transform canvas={scale=1.9},
  yshift=10, xshift=30,
  auto, node distance =0.8cm and 0.8cm, on grid, semithick,
  state/.style = {circle, top color=white, bottom color=white, 
                  draw, black, text=black, line width=0.1mm,
                  , scale=0.1mm},
  corner/.style = {rectangle, top color=white, bottom color=white, 
                  draw, white, text=black, scale=0.5},
]

\draw[pattern=north east lines, pattern color=black, line width=0.1mm] (1.2,0) -- (1.6, 0.692) -- (2, 0);

\node[state]  (Entry) at (1, 1.73) {};
\node[corner] (EntryNote) [right  = 1.3 of Entry] {~Program entry state};
\path[->, dashed, line width=0.1mm] (Entry) edge [] (EntryNote);

\node[state] (Fuzzing) [below right = 0.693 and 0.4 of Entry] {};
\node[corner] (FuzzingNote) [right  = 1.1 of Fuzzing] {Rare branch};
\path[->, dashed, line width=0.1mm] (Fuzzing) edge [] (FuzzingNote);

\node[state] (Dashed) [below right = 1.038 and 0.6 of Entry] {};
\node[state] (Node5) [below right = 1.038 and 0.2 of Entry] {};

\node[state] (Node1) [below left = 0.693 and 0.4 of Entry] {};
\node[state] (Node2) [below left = 1.039 and 0.6 of Entry] {};
\node[state] (Node3) [below left = 1.039 and 0.2 of Entry] {};
\node[corner, scale=1, fill=transparent] (Node4) [below = 0.8 of Node1] {\begin{tabular}{c} ... \\ log(n,m) \\ ...... \end{tabular}};

\coordinate  (Uncovered) at (1.6, 0.481);
\node[corner] (UncoveredNote) [right = 0.31 of Uncovered]
{\begin{tabular}{l}Unknown paths\end{tabular}};
\path[->, dashed, line width=0.1mm] (Uncovered) edge [] (UncoveredNote);

\coordinate  (Covered) at (1.2, 0.225);
\node[corner] (CoveredNote) [right  = 0.9 of Covered] {Observed paths};
\path[->, dashed, line width=0.1mm] (Covered) edge [] (CoveredNote);

\node[corner, scale=1] (Node6) [below right = 0.3 and 0.05 of Node5] {...};
\node[corner, scale=1] (Node7) [below left = 0.6 and 0.1 of Node5] {...};

\draw [decorate,decoration={brace,amplitude=3pt,mirror,raise=0pt},yshift=-1pt, line width=0.1mm] 
(0.801,0) -- (2, 0) node [black,midway,xshift=0pt, yshift=-10pt, scale=0.475] {Score: estimate the likelihood of finding new paths};

\draw [decorate,decoration={brace,amplitude=3pt,mirror,raise=0pt}, xshift=-3pt, yshift=0pt, line width=0.1mm] 
(1, 1.73) -- (0, 0) node [black,midway,xshift=-27.5pt, yshift=-18pt, scale=0.475, rotate=60] 
{\begin{tabular}{c} A concrete execution trace \\ of the common branch \end{tabular}} ;

\draw[line width=0.1mm, dashed] (0,0) -- (Node2);
\draw[line width=0.1mm] (Node1) -- (Node2);
\draw[line width=0.1mm] (Node1) -- (Node3);
\draw[line width=0.1mm, dashed] (Node4) -- (Node2);
\draw[line width=0.1mm, dashed] (Node4) -- (Node3);

\draw[line width=0.1mm] (Node1) -- (Entry) -- (Fuzzing);
\draw[line width=0.1mm] (0.801,0) -- (Node5) -- (Fuzzing) -- (Dashed);
\draw[line width=0.1mm] (1.2,0) -- (Dashed) -- (2, 0);

\end{tikzpicture}
  \caption{A tree representation of Ackermann02}
  \label{fig:AckermannInTree}
  \end{minipage}
  \vspace*{-.5cm}
\end{figure*}

% === Short version ===
% Constraint solving can compute inputs to penetrate the choke point (line~10) to reach the ``rare branch'' (lines~14/15),
% but then becomes unnecessarily expensive in solving the exponentially growing
% constraints from repeatedly unfolding the recursive function \texttt{ackermann}.
% By comparison, even though very few random fuzzer-generated inputs pass the chokepoint, the high speed of fuzzing means the ``rare branch'' will be quickly reached.
% === Short version ===

The program takes two inputs, \texttt{m} and \texttt{n} (lines~8/9), and then reaches a choke point in line~11,
which likely takes the ``common branch'' when \texttt{m} and \texttt{n} are chosen at random and thus immediately returns in line~12.

% Toby: cutting the following unless we can cite hard evidence for the claim
\iffalse
The snippet is a model of many modern softwares. 
For example, PDF readers need to validate the PDF file with a strictly restricted header, 
but allows almost infinite possibilities in the following content.
\fi

Concolic execution can compute inputs to penetrate the choke point to reach the ``rare branch'' (lines~14--16),
but generating sufficient inputs to cover all paths of the recursive function \texttt{ackermann} via constraint solving is unnecessarily expensive.
In comparison, a hypothetical random fuzzer restricted to generating values of $\texttt{m} \in \{0,1,2,3\}$ and $\texttt{n} \in \{0,\ldots,23\}$
will quickly uncover all paths of \texttt{ackermann} without the need to consider exponentially growing sets of constraints from the unfolding of its recursive calls.
Note that in general finding such closed-form solutions is expensive and/or undecidable for programs constraints that involve nonlinear arithmetic or bitwise operations.
However, we note that it is instead sufficient 
if the inputs preserve a chosen path prefix with high probability.
\iffalse
\FIXME{(did not understand the sentence)}.
\fi

Since they have complementary benefits, much prior work has sought to
combine concolic execution and fuzzing \cite{Driller, DigFuzz}.
But how should they be combined and when should each be used? For instance,
in the example
% when a hypothetical call to another function is inserted in the common branch at line~12 (cf. \texttt{log(m,n)}).
would it be more efficient to quickly flood both branches with unconstrained random inputs,
or to focus on uncovering the ``rare branch'' even though each input generation takes longer?

\Legion provides a general-purpose answer to this question by collecting and
leveraging statistics about the execution of the two branches.
\Legion collects statistics about program executions while simultaneously
iteratively exploring the program's execution paths and branching
structure, unifying this information together into a common tree-structured
search space on the fly. \cref{fig:AckermannInTree} illustrates such a
tree representation of \texttt{Ackermann02.c}. The tree root corresponds to the program entry point and every child node represents a conditional jump target
of its parent. Each node of the tree thus represents a partial program path.
Each stores statistics about the concrete executions observed so far that
follow that path (i.e.\ pass through that node).

At each iteration of the search,
these statistics allow \Legion to decide which node is most worthy of
further investigation. Having \emph{selected} a node, \Legion unifies
concolic execution (input-generation via solving path constraints)
\iffalse % cut this for now, not time to argue :)
\FIXME{not sure if it is better to say \textit{constraint solving} instead of \textit{concolic execution} at here, cause concolic often includes finding constraints, flipping constraints, and then solving for new paths. Here APPFuzzing only solves the given partial path constraint. To me, I consider the whole MCTS (which includes symbolic execution) to be concolic. Later in this paragraph, we also defined APPFuzzing again with solving+fuzzing}
\fi
and fuzzing in the form of \emph{approximate
  path-preserving fuzzing} (\APPF), to target that part of the
search space. 
\APPF is designed to \emph{sample} inputs that pass through the node,
i.e.\ to generate inputs that,  with high probability,
cause the program to follow the execution path from the root to the node
selected, but also distribute relatively randomly and uniformly among all
child paths of the selected node. \APPF uses a combination of
constraint solving, applied to the node's path condition, and controlled
mutation applied to solver-generated inputs.
Statistics are gathered from the concrete
executions produced by \APPF
to further refine \Legion's understanding of the search space of
the program under test, for subsequent iterations of the search.

In essence, these
statistics capture the value of sampling from a particular node. Since
\Legion's goal is to maximise coverage, its current implementation
measures value in terms of
\iffalse % I think we should put this detail in here. No need to bury it.
\FIXME{An alternative is to only say "coverage" at here + in the trade-offs (e.g. change "find new paths" to "increase coverage"), and leave the detail to \cref{sec: a reward evaluation heuristic}}
\fi
new execution paths uncovered (during concrete execution of the program
on those inputs).  Thus statistics such as the ratio of observed paths over total paths serve to estimate the potential of finding 
new paths by sampling from a particular subtree. This enables \Legion to address
the following two trade-offs:

\begin{enumerate}
  \item \textbf{The trade-off in depth} between a parent program state and its child. \label{ref: trade-off: depth}
  Sampling under the path constraints of a parent node (e.g.\ the choke point in the example) has three benefits: 
  a) Symbolic execution to the child states (common and rare branches) can be avoided; 
  b) inputs generated from the parent may traverse execution paths of all children; 
  and c) parents tend to have simpler constraints to solve. 
  However, this may waste time on children with constraints that are easy to satisfy.
  For instance, in the example, sampling the entry state of the program may cover both branches, but many executions will repeatedly traverse the common branch, 
  leaving the more interesting recursive function \texttt{ackermann} insufficiently tested. 
  Sampling from the rare branch state can guarantee executions to pass through the recursive function, 
  but it is more computationally expensive and will miss the function \texttt{log(n,m)}.
  
  \item \textbf{The trade-off in breath} between siblings program states. \label{ref: trade-off: breath}
  Given two siblings, sampling from either can potentially gain the benefit of covering more paths and collecting more statistical information of the program structure underneath the selected node, 
  at the opportunity cost of losing the same benefit on its sibling. 
   For example, choosing to sample inputs for the ``rare branch'' will cover more paths of the function \texttt{ackermann}, but it comes with a higher cost
   for input generation via constraint solving and can neither cover the
   sibling nor learn about the subtree beneath it.
\end{enumerate}

\Legion's statistical approach allows it to address these trade-offs by
adopting neither a breath- nor depth-first approach, but instead
a \emph{best-first} strategy. 
% false

\Legion's best-first strategy is a variation of the Monte Carlo tree search algorithm~\cite{MCTS-Survey}, a popular AI search strategy for large search problems in the absence of full domain knowledge. A sequential decision-making framework, MCTS carefully balances the choice between selecting
the strategy that appears to be most rewarding based on current information (\emph{exploitation}),
vs one that appears to be suboptimal after few observations but may turn out to be superior
in the long run (\emph{exploration}).

We argue that MCTS is particularly well-suited for automated coverage testing,
not only due to its utility on large search spaces without full domain
knowledge or even domain heuristics, but also because it is known to perform well on asymmetric search
spaces (as exhibited in the running example and ubiquitous in software generally).

To utilise the full potential of MCTS on coverage testing, \Legion
addresses the following challenges:
\begin{itemize}
\item The selection policy of MCTS, controlled by its \emph{score} function
  (explained later in \cref{sec: selection policy}),
  plays a key role in its performance. However there is no well-established strategy or heuristic in coverage testing to determine the most efficient program compartment to focus in different scenarios.
  Thus \Legion implements a \emph{modular} score function, flexible to
  a range of heuristics (\cref{sec: selection policy}) that we show can be
  efficiently instantiated for coverage-based testing (\cref{sec: a reward evaluation heuristic}).
\item How should MCTS
  hyperparameters (\cref{sec: hyper-parameters}) be
  chosen for a program under test, and what is their influence on
  \Legion's performance? We answer this question by empirical
  evaluation, showing that \Legion can perform effectively with naive
  hyperparameter choices across a wide variety of programs, as well
  measuring the sensitivity of its performance to the choice of parameters
  (\cref{sec:evaluation}).
\item To achieve maximal efficiency MCTS requires being able to randomly
  and uniformly simulate
  plays from the node selected at each iteration.  For \Legion this means
  being able to generate many random and uniform inputs that can traverse the selected program state by mutating solutions from constraint solving. Neither fuzzing nor constraint solving alone are
  sufficient for this purpose, for which \Legion introduces approximate
  path-preserving fuzzing, explained in \cref{sec: APPFuzzing algorithm}.
\end{itemize}

%MCTS also requires accurate tracing for reward evaluation and propagation (Sec. \cref{sec: tracejump}).
% For example, after seeing $A \leftarrow B$, $B \leftarrow C$, and $C \leftarrow D$, 
% AFL will not consider trace $A \leftarrow B \leftarrow C \leftarrow D$ as an interesting new sub-path, 
% causing to miss a new path. This is addressed by \cref{sec: tracejump}

\section{\Legion MCTS} \label{sec: Legion MCTS}
\iffalse % Toby: cut this
{\color{orange}
1. Higher coverage
2. Flexible exploration strategy
3. Generalise of concolic execution and fuzzing
4. Learn from the statistic
5. Can analysis which technique is more suitable for which types of programs.

}
\fi

The key insight of \Legion is to generalise concolic execution and fuzzing in a principled way with the Monte Carlo tree search algorithm.
% Toby: don't end paragraph here; comment lines prevent LaTeX from doing so
% 
%  ==== Vanilla Monte Carlo tree search ====
% The Vanilla MCTS formulates search space exploration under uncertainty into a sequential decision-making process and 
% constructs a search tree that represents part of the whole search space in the following four steps:
%
% \begin{itemize}
%   \item Selection. 
%   It descends the current search tree according to a selection policy 
%   by selecting the node with the highest score and stops at an expandable node. 
%   \item Expansion. 
%   It expands the tree with the children of the node selected for one step.
%   \item Simulation. 
%   It randomly simulates from the new leaf node to the end in order to produce an outcome.
%   \item Back-propagation. 
%   It back-propagates the simulation result to all nodes along the selection path to update their scores. 
%   This result will accumulatively contribute to the selection of the next iteration.
% \end{itemize}
%
% ==== Vanilla Monte Carlo tree search ====
To do so,
\Legion makes two modifications to traditional MCTS to make it more suitable for coverage-based testing, which we describe in this section. These
concern respectively the MCTS tree nodes and the tree construction.

\subsection{Tree Nodes} \label{sec: tree nodes}

As mentioned in \cref{sec: Overview}, \Legion treats testing as searching
a tree-structured space that represents the reachable states of the program,
an exemplar of which is depicted in \cref{fig:AckermannInTree}.
Each tree node is identified by the address of a code block in the program under test. The tree root corresponds to the program entry and every child node represents a conditional jump target of its parent.

As mentioned, each node represents the (partial) execution path from the program's
entry point to it, and stores statistics about the concrete program executions
observed so far to follow that path, which are used by the MCTS selection
policy to decide which part of the tree to investigate at each iteration
of the search algorithm. When the MCTS algorithm selects a node for
investigation, \Legion then generates new program inputs that (with
high probability)
cause the program to traverse the path represented by that node. To do so,
recall that \Legion
uses a form of directed fuzzing called approximate path-preserving fuzzing
(\APPF), which is a hybrid of mutation fuzzing and input-generation by
constraint solving. \Legion's \APPF implementation thus benefits from
symbolic information about the program state that corresponds to the tree
node (i.e.\ the state reached after traversing the partial path that the
node represents), specifically the symbolic path
condition. Therefore, certain nodes in the tree also carry a
\emph{symbolic state}, which encodes the corresponding path condition.

However, not all tree nodes carry a symbolic state. Recall, from
\cref{sec: Introduction} that \Legion performs symbolic execution lazily.
This means that after observing a new
concrete execution path, that path will be integrated into the tree but
without symbolically executing it. Indeed \Legion defers symbolically
executing a path until it has evidence that investigating that path could
be beneficial for uncovering additional new execution paths (as determined by
the MCTS selection algorithm). Until a path is symbolically executed, the
tree nodes that represent it do not contain symbolic state information.

\begin{figure*}[ht!]
  \centering
  \input{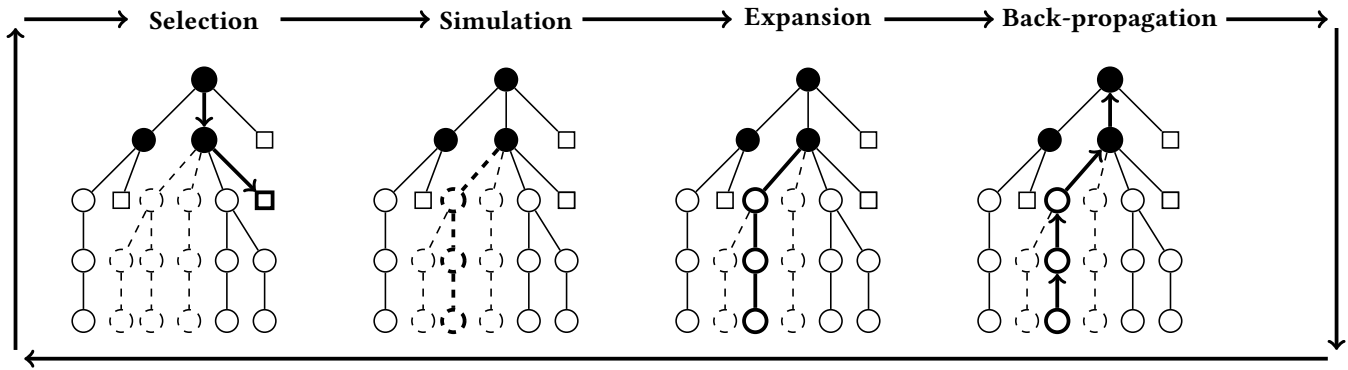}
  \vspace*{-.3cm}
  \caption{The four stages of (each iteration of) \Legion's MCTS algorithm.}
  \label{fig: legion phases}
\end{figure*}

Thus, \Legion's tree nodes come in a variety of types, depending
on their purpose and the information they contain. For instance,
hollow nodes are ones that represent observed concrete execution paths
but do not (yet)
carry symbolic state information, while solid nodes additionally
do carry satisfiable symbolic state information; 
phantom nodes represent feasible program paths
observed and validated during symbolic execution but not yet observed during concrete
execution. Two further types also exist: redundant and simulation,
as described below. We discuss each type of node, with occasional
reference to
\cref{fig: legion phases}.\prebparagraph

\bparagraph{Hollow nodes} are basic blocks found by concrete, binary
  execution, but have not been selected for symbolic execution yet, and
  hence do not have their symbolic states attached. They are used to mark the existence (and reachability) of paths and to collect statistics of observed rewards. When a hollow node is selected for investigation by the MCTS selection step (explained shortly), \Legion invokes symbolic execution from its closest solid ancestor, to obtain a path condition to seed \APPF. By doing so, the
  node will be re-categorised as one of the following two types. Each hollow node appears as a hollow round node in \cref{fig: legion phases}.\prebparagraph

\bparagraph{Solid nodes} are hollow nodes with symbolic states attached and whose path condition has one more constraint over that of its closest solid ancestor, i.e.\ they are not redundant. Each solid node has a special child node called
  a \emph{simulation child}  (of square shape), described below. Each solid node appears as a solid round node in \cref{fig: legion phases}.\prebparagraph

\bparagraph{Redundant nodes} represent states for which \APPF is (a)~impossible or (b)~redundant: either (a)~they were observed during concrete execution but not during symbolic execution (e.g.\ due to under-approximation and concretisation during symbolic execution), and so cannot be selected for \APPF because they lack symbolic state information,  or (b)~have exactly the same symbolic constraint as their closest solid parent (e.g.\ jump targets of tautology conditions), for which \APPF is redundant because identical information is already captured by their closest solid ancestor. Hence \Legion never simulates from them (i.e.\ never selects them for \APPF). No redundant nodes are depicted in \cref{fig: legion phases}.\prebparagraph
  
  \bparagraph{Phantom nodes} denote symbolic states whose addresses have not yet been seen during concrete execution, but have been proved to exist by the symbolic execution engine. They are found during the selection stage, when the symbolic execution engine shows that the symbolic state of a hollow lone child has a sibling state.   
    They act as place holders for as-yet unseen paths, waiting to be
    revealed by concrete, binary execution (i.e.\ by \APPF).
    When that happens, a phantom node will be replaced by a solid node, when
    \Legion integrates the observed concrete execution path into the tree.
    No phantom nodes are depicted in \cref{fig: legion phases}.\prebparagraph

  \bparagraph{Simulation nodes.} Different from vanilla MCTS, \Legion allows sampling from intermediate nodes, which represents \APPF applied to some point along a partial execution path. To incorporate this into MCTS, we add special leaf nodes to
    the tree whose shape is a square and whose parent is always solid.
    When the MCTS selection stage chooses a simulation node for \APPF, this
    choice implies that \Legion believes that directing fuzzing towards the
    program state represented by the solid parent is more beneficial than fuzzing any of its child program states (where ``beneficial'' means the estimated
    likelihood of discovering a new path). Each simulation node appears as a square node in \cref{fig: legion phases}.

\subsection{Tree Construction}

\cref{fig: legion phases} illustrates how \Legion 
uncovers the tree-structured search space in its variation of MCTS:
Each iteration of the search algorithm proceeds in four stages.
In  \cref{fig: legion phases}, bold lines highlight the actions in each stage. 
Solid thin lines and nodes represent paths and code blocks that have been
covered and integrated into the tree.
Dotted thin lines and nodes represent paths and nodes not yet found.
The four stages are as follows: \\[-0.8em]

\bparagraph{Selection.} 
  Shown by bold arrows, \Legion descends from the root by recursively applying a \textit{selection policy} (\cref{sec: selection policy}), %which guides it towards the part of the tree deemed most worthwhile to investigate, 
  until it reaches a simulation node (square shape in \cref{fig: legion phases}). 
  Upon visiting a hollow node (i.e.\ that carries no symbolic state,
  depicted as hollow circles),
  it performs symbolic execution from the nearest solid ancestor
  (depicted as filled nodes) to compute the
  symbolic state of the hollow node, adds the
  symbolic state to that node (turning it solid) and attaches to it a
  simulation child (the square).\\[-0.8em]

\bparagraph{Simulation.} Having selected which program state to target
  for investigation, \Legion applies \APPF (\cref{sec: APPFuzzing algorithm}) on the program state represented by the (solid parent of) the selected simulation node to generate a collection of inputs that, with high probability, will cause the program to follow the path from the root to the solid parent.
  It then executes the program on those inputs and observes the resulting
  execution traces, which are represented by dashed circles and lines in
  \cref{fig: legion phases}. \\[-0.8em]

\bparagraph{Expansion.} This step involves mapping each observed
  execution trace to a path of corresponding tree nodes: new traces cause
  new hollow (round) nodes to be added to the tree.\\[-0.8em]

\bparagraph{Back-propagation.} This step updates the statistical
  information recorded in the tree, to take account of the executions observed
  during the simulation step. \Legion computes the \emph{reward} (\cref{sec: a reward evaluation heuristic}) from
  each simulation (i.e.\ from each concrete execution performed during the simulation step)
  and propagates the reward to all nodes along the execution trace.
  Note that the path(s) observed during concrete execution might differ from
  that chosen during the simulation step, because \APPF is necessarily
  approximate. Therefore, in this step, \Legion also propagates the reward
  to each node on the path from the root to the node chosen during the
  selection step.
  % === An alternative way to describe the 4 stages ===
  % \item It first starts with the simulation step, where we execute the instrumented program binary with certain inputs. In particular, we use the seeded inputs in the first round and mutated inputs in the following iterations. Mutated inputs are generated by \hyperref[sec: PPF]{\textit{path preserving fuzzing}}, which ensures the inputs will traverse the target program location to explore. Each execution will generate a traces of addresses representing the basic blocks marked by the instrumentation.
  % \item It then follows with the expansion step, which integrates these new execution traces to the existing tree. Each address in the trace maps to a tree node, therefore each trace can be uniquely identified by the array of addresses from the root node to any leaf node. During the expansion process, it evaluates the outcome of this iteration based on the number of new paths found and the tree nodes traversed.
  % \item The third step propagates the outcome to all related nodes and update their scores accordingly. The score of each node measures the potential of a node regarding discovering more unrevealed paths from it.  
  % \item The final step is selection, which identifies the program location that has the highest score.
  % === An alternative way to describe the 4 stages ===

\iffalse % Toby: cut orange stuff
{\color{orange}
  Four stages of \Legion MCTS}
\fi

\section{Design Considerations} \label{sec: Design Considerations}

Having described \Legion's adaptation of MCTS at a high-level,
we now discuss two of the most critical parts of its design, namely the
design of the selection policy used during the selection phase of each MCTS
iteration, and the design of \Legion's directed fuzzing implementation,
\APPF. We discuss further implementation details and considerations 
in \cref{sec: practical considerations}.

\subsection{Selection Policy} \label{sec: selection policy}

One of the central design goals of \Legion is to generalise a range of prior
coverage-based testing methods. Doing so requires a general policy for
selecting which tree node to investigate at each MCTS iteration,
to make \Legion adaptable to different search strategies and coverage
reward heuristics, with a modularised design.

At its simplest, the selection policy's job it to assign a \emph{score} to
each node. This score is used as follows, during the selection step of
each MCTS iteration. Recall that selection proceeds via recursive descent
through the tree towards that part of the tree deemed most worthwhile to
investigate. The score guides this descent. Selection starts at the root.
Then the immediate child with the highest score is traversed
(with ties being broken via uniform random selection). Traversal
proceeds to that child's highest-scoring child, and so on.

\newcommand{\UCT}{\mathit{UCT}}
\newcommand{\Nsel}{N_{\mathit{sel}}}
\newcommand{\Psel}{P_{\mathit{sel}}}
In \Legion, the score of each node represents the optimistic estimation that
\APPF will discover new paths when applied to the (program state represented by the) node.
Scores are computed by applying the \textit{Upper Confidence Tree (UCT) algorithm}~\cite{MCTS-Survey} on the rewards obtained during the prior simulation (i.e.\ during the prior \APPF). We leave a
discussion of rewards to \cref{sec: a reward evaluation heuristic} but for now
it suffices to understand that rewards correspond to the discovery of new
execution paths.
For a node~$N$ we define its score, $UCT(N)$ as follows. 
For a newly created node, its score is initialised to $\infty$, to ensure that uninvestigated subtrees are always prioritised over their siblings. Otherwise, $UCT(N)$ is:
\begin{equation}\label{eqn:UCT-general}
\UCT(N) =  \bar{X}_N + \rho\sqrt{\frac{2\ln{\Psel}}{{\Nsel}}}
\end{equation}
where $\bar{X}_N$ is the average past reward from~$N$, 
$\rho$ is a hyperparameter (described below) that defines the MCTS \emph{exploration ratio}, $\Nsel$ and $\Psel$ are, respectively,
the number of times that node~$N$ and its parent have been traversed
so far during prior selection stages. % \FIXME{check for consistent notation with later in the paper.}

This score is designed to balance two competing concerns when searching
the tree, namely
\emph{exploitation} vs.\ \emph{exploration}. Exploitation corresponds to
investigating a part of the tree that has proved rewarding in the past (high $\bar{X}_N$),
in the belief that because it has yielded new execution paths before it is
likely to do so again in the future. Exploration, on the other hand,
correspond to investigating a part of the tree that, so far, appears
under-explored (low $\Psel/\Nsel$), and where rewards are less certain. The second term derives from an upper bound of a confidence interval for the true mean, based on the multi-armed bandit algorithm UCB1~\cite{kocsis2006bandit}.

The precise balance between these two concerns is controlled by the choice
of non-negative
hyperparameter~$\rho$. Were $\rho$ zero (or very large), \Legion would
base its selection decisions only on past rewards (resp. visitation counts), corresponding to pure exploitation (resp. exploration).
Instead $\rho$ should be chosen to be some small positive
value, to permit exploration. We investigate in more detail
how $\rho$ affects \Legion's performance. For any fixed choice, as a node's parent is traversed without the node visited, the fraction $\sqrt{\frac{2\ln{\Psel}}{\Nsel}}$ grows, causing \Legion to favour the child. %This corresponds to exploration: choosing to investigate a part of the tree that appears under-investigated (relative to its ancestors).

%The various constants that appear in the score function are standard. \FIXME{citation, further justification?}

\subsection{Approximate Path-Preserving Fuzzing} \label{sec: APPFuzzing algorithm}

Having selected a node (i.e.\ program state) to target, \Legion then applies
its approximate path-preserving fuzzing algorithm \APPF to generate inputs that
when supplied to the program under test will,
with high probability, cause the program to execute from
the entry point to reach that state, following the path from the tree root
to the selected node.

\APPF can be seen as a hybrid of input generation via constraint solving
and mutation fuzzing, and is seeded by the
symbolic path condition stored in the selected simulation node. 
\Legion's \APPF is inspired by~\QuickSampler \cite{QuickSampler}, a recent
algorithm for sampling likely solutions to Boolean constraints that mixes
SAT solving and bit mutation. \Legion's \APPF on the other hand operates on
SMT constraints, specifically the bit-vectors theory of the constraints
produced by \Legion's symbolic execution engine, \Angr~\cite{angr}.

\newcommand{\constraint}{\mathit{constraint}}
\newcommand{\Nsamples}{N_{\mathit{samples}}}

\begin{algorithm}[h!]
  \DontPrintSemicolon
  \SetKwFunction{APPFuzzGen}{APPFuzzGen}%
  \SetKwProg{Fn}{def}{:}{}
  \Fn{\APPFuzzGen{$\constraint$}}{

    $\Sigma = \{\} $\;
    $\sigma = \mathit{solve}(\constraint)$\;
    \textbf{yield} $\{\sigma\}$ \;

    % $mutants \leftarrow \sigma$\;
    \For{each bit $b_i$ of $\sigma$} {
      $\tilde{\Sigma} = \{\}$\;
      $\sigma' = \mathit{solve}(\constraint \land \lnot b_i)$\;
      $\tilde{\Sigma} = \tilde{\Sigma} \union \{\sigma'\}$\;
      \For{$\sigma''$ in $\Sigma$}{
        $\tilde{\Sigma}  = \tilde{\Sigma} \union \{(\sigma \oplus ((\sigma \oplus \sigma') \vee (\sigma \oplus \sigma''))\}$\;
      }
      \textbf{yield} $\tilde{\Sigma}$ \;
      $\Sigma = \Sigma \cup \tilde{\Sigma}$      
    }    
  }
  \caption{Input Generation for \APPF\label{algorithm: APPFuzzGen}}
\end{algorithm}

\begin{algorithm}[h!]
  \DontPrintSemicolon
  \SetKwFunction{APPFuzz}{APPFuzz}%
  \SetKwProg{Fn}{def}{:}{}
  \Fn{\APPFuzz{$\Nsamples$, node}}{
    results = []\;
    % $mutants \leftarrow \sigma$\;
    \While{len(results) < $\Nsamples$} {
      results.append(\APPFuzzGen(\emph{node.path\_constraint}))\;
    }    
  }
  \caption{Approximate Path-Preserving Fuzzing \label{algorithm: APPFuzz}}
\end{algorithm}

The \APPF algorithm, \APPFuzz, is depicted in \cref{algorithm: APPFuzz}. Its inputs comprise the selected node to which \APPF is being applied, as well
as constant $\Nsamples$ the \emph{minimum} number of new inputs
that \APPF should attempt to generate. Like $\rho$,
$\Nsamples$ is a hyperparameter, discussed further in \cref{param: sample number}.

Input generation is delegated to
a helper \emph{generator} function, depicted in \cref{algorithm: APPFuzzGen}, which is 
repeatedly called until at least $\Nsamples$ have been generated.
By design, the helper generator, \APPFuzzGen, can yield a variable number of
inputs each time it is
called. Hence, \APPFuzz can often end up generating more than the minimal
number of required inputs for each simulation.

Input generation, handled by \APPFuzzGen in \cref{algorithm: APPFuzzGen},
is performed using a mixture of constraint solving and mutation of prior
solver-generated solutions. Thus \APPFuzzGen takes as a parameter the
path constraint~$\constraint$ of the node in question. On its first invocation
it simply solves the node's constraint and returns that solution~$\sigma$.
However, when subsequently invoked for the same node
(e.g.\ if multiple inputs are required
for simulation,
when $\Nsamples > 1$) its execution restarts from the point directly after
the first {\bf yield} statement, i.e.\ at the entry to the outer loop.
Each iteration of this
outer loop generates an ever-larger set of new inputs, and the generator
returns the newly-generated inputs after each iteration. Subsequent
invocations of the generator
resume execution from the point just after the second {\bf yield} statement,
i.e.\ at the point where the next iteration of the outer-loop proceeds.

The outer loop iterates over each bit~$b_i$ of the initial solver-generated
solution~$\sigma$. 
In each outer loop iteration, the generator
 invokes the
 constraint solver once to generate a solution~$\sigma'$ to the path constraint that,
 if one exists,
 is guaranteed to be distinct from~$\sigma$ by differing in bit~$b_i$.
 Then, for each input~$\sigma''$
 generated in previous iterations of the outer loop, $\sigma'$ is \emph{mutated}
 with $\sigma''$ and~$\sigma$ to produce a new input. The $\sigma'$ and
 all of the new mutation-generated inputs are then returned.

 Thus each invocation of the generator causes a single invocation of the
 constraint solver; however repeated invocations yield exponentially more
 inputs. This process continues until the outer-loop terminates.
 The generator will begin afresh on subsequent invocations.

 Not depicted in \cref{algorithm: APPFuzzGen} is the fact that previously generated
 inputs are remembered, to avoid returning duplicate inputs. Once the solver
 can return no new solutions, a node is marked as \emph{exhausted} and no
 further simulation (\APPF) will be performed on it (see \cref{sec:optimisations}).

 The mutation operator combining $\sigma$, $\sigma'$ and $\sigma''$, via
 bitwise exclusive-or $\oplus$ simulates the mutation operator used in a
 similar fashion by \QuickSampler~\cite{QuickSampler}
 on Boolean constraints, lifting it to bit-vectors. 
   That operator was shown, with high probability, to produce results that each
 satisfy the Boolean constraint~\cite{QuickSampler} and, in aggregate,
 are uniformly distributed.

 For this reason, we conjecture that our \APPF implementation is likely to
 produce inputs that preserve the path to the selected node and that
 satisfy i.i.d. requirements of MCTS simulations.

 In scenarios where one wishes to produce more inputs per constraint solution,
 our implementation can be adjusted to do so by having it perform more
 mutations per outer loop iteration. However doing so is likely
 to reduce the accuracy of the results, i.e.\ the probability that they will
 satisfy the given path condition.

 \iffalse % Toby: i suggest we cut this for now to save space, since it is
 % a detail that all concolic executions need to implement and so is not
 % a contribution
\subsection{TraceJump} \label{sec: tracejump}

Accurate concrete execution tracing of the inputs from APPF is achieved with a tool named \TraceJump. 
It instruments the source code to record all conditional jump targets encountered during each concrete execution. 
Each jump target is identified by the first address of its code block. 
During the expansion stage, each sequence of addresses will be integrated into the search tree as a line of tree nodes. 
Simulation rewards will be computed from these traces and propagate back to corresponding nodes.
These addresses will also be used to pair tree nodes with symbolic states in the selection stage when a tree node is selected for the first time.

\TraceJump logs addresses in the \texttt{stderr} file descriptor. It also buffers addresses to amortise the cost of \texttt{syscall}.
\fi

\section{Practical Considerations} \label{sec: practical considerations}

With the major design considerations out of the way, we now turn to the
salient details of \Legion's current implementation. These concern
respectively the reward heuristic used in the UCT score function of its
MCTS implementation, as well as the choices of the various hyperparameters
like $\rho$ and $\Nsamples$.

\subsection{A Reward Evaluation Heuristic} \label{sec: a reward evaluation heuristic}

In MCTS, rewards quantify the benefits observed during each simulation.
In \Legion's MCTS, simulation corresponds to concrete execution of the
program under test, via \APPF.
An important implementation consideration is what should be the reward
function? Put another way, what should rewards correspond to? When should
a simulation (concrete execution) be considered rewarding?

Recall that the average past rewards associated with a node~$N$ was denoted
$\bar{X}_N$, in the UCT score function defined in \cref{eqn:UCT-general}
in \cref{sec: selection policy}.
By choosing different instantiations for this term, one can naturally
adapt \Legion to various exploitation strategies.

However, since the goal of \Legion's present implementation is to
discover the maximum number of execution paths in the shortest possible time,
in this paper we implement and evaluate a simple reward heuristic:
the reward of a simulation is the number of new paths found by \APPF.

\newcommand{\Nwin}{N_\mathit{win}}
Recall that for a node~$N$, $\Nsel$ denotes the number of times~$N$ has
been traversed during selection, and that $\Psel$ does likewise for~$N$'s
parent. Then we denote by $\Nwin$ the number of distinct execution paths
discovered so far that pass through node~$N$.

The average past reward associated with node~$N$ is then $\frac{\Nwin}{\Nsel}$,
the ratio of the number of paths found so far that pass through this node,
compared to the number of times it (or a child) has been selected for \APPF.
This ratio is chosen under the assumption that
the more new paths were found by running \APPF on a node in the past,
the more potential the node has in increasing coverage in the future.

Plugging this into \cref{eqn:UCT-general}, and remembering that the initial
score of a node is set to~$\infty$, we arrive at the following instantiation
of the UCT score in \Legion's current implementation:

\begin{gather*}
  \UCT(N) =  
  \begin{cases}
    \infty  & \Nsel=0\\
    \frac{\Nwin}{\Nsel} + \rho\sqrt{ \frac{2\ln{\Psel}}{\Nsel} } & \Nsel > 0
  \end{cases}
\end{gather*}

% But this heuristic can be replaced with others that reflect some statistics of concrete executions (concrete execution time, constraint solving time, etc.) 
% or tree structure (depth of the node, etc.). More in Sec. \cref{sec: future work}.
To better understand this heuristic, let us return to the example of
\cref{lst:ackermann02}. 
Note that only a specific pair of \texttt{m} and \texttt{n} can violate the assertion on line 15, but uncovering that assertion requires being able to get past the recursion of the \texttt{ackermann} function. The recursion of
\texttt{ackermann} is therefore an attractive nuisance for uncovering the
\texttt{assert} statement, since recursion necessarily produces
new execution paths. This is why the exploration component of the UCT score
is critical. At the same time, the use of such a fine-grained coverage
metric, which tracks individual execution paths, instead of more coarse-grained
metrics like statement and branch coverage or AFL's coverage maps, ensures
that \Legion does not overlook paths that can arise only after
deep recursion.

As we show later in \cref{sec: experiment results}, even with this
simple heuristic, \Legion performs surprisingly well, and the heuristic
appears relatively robust. Yet there of course exists much scope to consider
heuristics that take into account other relevant information, such as
time consumption, subtree size or other static properties of the program.
We leave the investigation of such for future work.

\subsubsection{Optimisations}\label{sec:optimisations}

This heuristic permits a number of optimisations on node selection,
which we have implemented in \Legion. Essentially these allow \Legion to decide when a
node will no longer produce any future rewards (i.e.\ that no new paths can be
found via \APPF applied to the node). When doing so, \Legion overrides the
node's score to $-\infty$. In this case we say that the node is
\emph{pruned} from the tree. However note that the node is never physically
removed from the tree: its score is just overridden to ensure it will never
be selected.

\textbf{Fully explored nodes.} A node is fully explored if there is no undiscovered path beneath it. For example, final nodes of complete concrete execution
paths are fully explored.
A parent node will be pruned if all non-simulation children are fully explored.
Given a fully explored node may have hidden siblings, \Legion will prune it only after identifying all of its siblings via symbolic execution.

\textbf{Useless simulation nodes.} A simulation node is useless if it has less than two not fully explored siblings. Recall that simulation nodes are children of (solid) nodes and that solid nodes represent a point in a concrete execution path and contain a symbolic path condition. When a simulation node is selected during the MCTS selection stage (\cref{sec: Legion MCTS}) this represents the decision to target its solid parent node with \APPF. Doing so essentially will yield some paths from the full set of paths~$S$ in the subtree under the solid parent. On the other hand, sampling from one of the simulation node's siblings
(i.e.\ from another child of the solid parent) will yield paths that are drawn from a strict subset~$T$ of~$S$. Sampling from~$S$ can be beneficial if it can yield paths from multiple such~$T$. However,
if there is only one such~$T$ remaining, it is better to sample from it than from~$S$. Hence, in this case, the simulation node is pruned.

\textbf{Exhausted simulation nodes.} A simulation node is exhausted if no input can be found when applying \APPF on it. This happens when all inputs that the solver is capable of producing under the constraint of the node have been generated.

\textbf{Under-approximate symbolic execution.} Due to under-{\linebreak}approximation, the symbolic execution engine might incorrectly classify a feasible state as infeasible, e.g. due to early concretisation. This creates a mismatch between the symbolic execution results and concrete execution traces. In this case
\Legion cannot apply \APPF to such a node or any of its decedents, but must instead target its parents. Under-approximation can cause \Legion to erroneously conclude that a subtree has been fully explored. To mitigate this issue, \Legion can be run in ``persistent'' mode, which continues input generation even for apparently fully-explored subtrees (see \cref{sec: peer competition settings}).

% \textbf{A subtree with any uncovered path.} Under this heuristic, \Legion only selects the nodes that may have undiscovered paths.
% The uncertainty of this heuristic only comes from the potential of finding new paths.
% The symbolic execution engine guarantees to uncovered all program states and hence clear up such uncertainty.

% \textbf{Leaf nodes.} Here we only consider non-simulation nodes.
% Non-hollow leaf nodes will be pruned from the selection stage. Leaf tree nodes are the addresses of the last conditional jump targets captured by \TraceJump, they don't have any subsequent address to be uncovered. They might have sibling nodes, which will be uncovered by symbolic execution after they are dyed. A leaf node is considered fully explored when it is dyed.

% \textbf{Root of a fully explored subtree.} A node and its subtree will be pruned from the selection stage in two cases: 
% \begin{itemize}
%     \item All nodes in its subtree are fully explored.  After all nodes are confirmed to be fully explored, the subtree has no new paths to discover. 
%     \item No inputs can be found by \APPF on its simulation node. The exhaustion of \APPF implies all inputs under the constraint of the node have been generated and executed, hence all feasible paths are found.
% \end{itemize}
\ifnew
Recall from \cref{sec: Overview} that to construct the search tree
\Legion uses binary instrumentation to
collect the addresses of conditional jump targets traversed during concrete execution. Doing so produces a trace (i.e.\ a list)
of conditional jump target addresses.
\Legion can be configured to limit the length of such traces to a fixed bound,
mitigating the overheads of instrumented binary execution~\cite{VeriFuzz} and,
hence, the depth of the search tree.
The number is controlled by a hyperparameter named \emph{tree depth} (see \cref{param: tree depth}).
\fi

\subsection{Hyperparameters} \label{sec: hyper-parameters}

Another important practical consideration is the choice of the various
hyperparameters that control \Legion's behaviour. Some of these, like
$\rho$ and $\Nsamples$ we have already encountered. However we summarise
them all below for completeness.

Naturally, different choices of
the hyperparameters will bias \Legion's performance in favour of certain
kinds of programs under test. We investigate this effect in
\cref{sec: sensitivity experiment results}.

\iffalse % cut for space by Toby. it is not really a hyperparamter i think
\textbf{Score function.} \label{param: score function}
\Legion's score function computes the score of each tree node.
\cref{sec: senstitivity experiement settings} compares our heuristic
(discussed in \cref{sec: a reward evaluation heuristic})
against a baseline that assigns a random value to each node.
\fi

\textbf{Exploration ratio $\rho$.} \label{param: exploration ratio}
The exploration ratio of the MCTS algorithm controls the amount of exploration performed, in comparison to exploitation (see \cref{sec: selection policy}), by the selection phase of MCTS.

\ifnew
\textbf{Number of cores.} \label{param: number of cores}
\Legion supports the \emph{leaf parallelisation} of MCTS, wherein simulations
are run in parallel. For \Legion this corresponds to running multiple
concrete executions in parallel, when \APPF returns multiple inputs.
The maximum number of such simultaneous parallel executions is controlled
by this hyperparameter.

\textbf{Tree depth.} \label{param: tree depth}
\Legion can limit the depth of its search tree to this parameter by forcibly
terminating concrete executions once the length of the
trace of conditional jump targets produced by the binary instrumentation
reaches this value.
\fi

\iffalse % Toby cutting this as it raises more questions than it answers
Empirically, this early termination mitigates memory leakage caused by overlong instrumented concrete executions via \TraceJump when launched as a subprocess of \Python. The reason to the memory leakage is cunrrently unknown.
\fi
% Long instrumented concrete executions via \TraceJump launched as a subprocess of \Python tends to cause memory leakage for currently .
% As a consequence, we ran into a lot of out-of-memory situations much earlier than the timeout of~900s in the competition.
% We mitigate this issue by only tracing upto certain number of addresses in each execution.
% The number is controlled by a hyperparameter \textit{tree depth} \cref{param: tree depth} and we proved that it does not largely affect performance of \Legion when chosen in a reasonable range.

\textbf{Concrete execution timeout.} \label{param: conex-timeout}
This hyperparameter controls when \Legion will forcibly terminate
concrete executions (produced by \APPF) that take too long. 

\textbf{Symbolic execution timeout.} \label{param: symex-timeout}
Recall that when \Legion decides to target 
a  node for \APPF that does not yet contain its symbolic path condition,
it then symbolically executes to that node from its nearest symbolic
ancestor. The symbolic execution
timeout is used to forcibly terminate such symbolic executions if they
take too long. If such a timeout occurs, \Legion will proceed to
select the nearest symbolic ancestor produced by the terminated symbolic execution.

\textbf{Minimum number of samples per simulation~$\Nsamples$.} \label{param: sample number}
Recall that this parameter controls the minimum number of inputs to
generate for each \APPF invocation, and that each such invocation might
return more than this minimum (see \cref{sec: APPFuzzing algorithm}).

\textbf{Persistent.} \label{param: persistent}
When set, this boolean flag causes \Legion to continue input generation even for apparently fully-explored subtrees to mitigate under-approximate symbolic execution (\cref{sec:optimisations}) and expensive constraint solving.

\iffalse % Toby: cut these since they are not mentioned again later

\textbf{Coverage-only.} \label{param: coverage-only}
\Legion will aim for maximum coverage and will not terminate after identifying bugs.
This flag is used in all experiments in this paper.
\todo{1.6}

\textbf{Number of cores}.
\Legion supports the \emph{leaf parallelisation} of MCTS, wherein simulations
are run in parallel. For \Legion this corresponds to running multiple
concrete executions in parallel, when \APPF returns multiple inputs.
The maximum number of such simultaneous parallel executions is controlled
by this hyperparameter.
% For the experiments performed in this paper,
% we set this value to $8$ via the flag \texttt{--core 8}.

\textbf{Maximum input length}.
When selecting the root node for \APPF, \Legion must generate an input
for it. Since, by definition, the root node has the trivial path constraint,
\Legion simply employs random generation. This hyperparameter controls
the maximum length random input that will be generated in this case.
\Legion currently assumes all inputs are within $1000$ bytes.
\fi

% Toby: cut
\iffalse
{\color{blue} Useless flags, e.g. time penalty (which is 0)}
\fi

\newcommand\Tstrut{\rule{0pt}{2.4ex}}       % "top" strut
\newcommand\Bstrut{\rule[-0.9ex]{0pt}{0pt}} % "bottom" strut
\newcommand{\TBstrut}{\Tstrut\Bstrut} % top&bottom struts

\begin{table*}[!ht]

\resizebox{\textwidth}{!}{\begin{tabular}{|r@{ = } l|c|c|c|c|c|c|c||c|S[table-format=1.10, detect-weight,mode=text]|}
    \hline
    \multicolumn{2}{|r|}{Settings} & Score function & $\rho$ & Core & Tree Depth & ConEx Timeout & SymEx Timeout & $\Nsamples$ & Score & {p-value} \\
    \hline\hline
    \multicolumn{2}{|r|}{Baseline} & UCT  & $\sqrt{2}$  & $8$ & $10^5$ & $-$ & $-$ & $1$ & $48.9018$ &  \TBstrut\\
    \hline\hline
    Score Function & Random  & \textbf{Random} & $\sqrt{2}$  & $8$ & $10^5$ & $-$ & $-$ & $1$ & $48.1632$ &\bfseries 0.004404148 \TBstrut\\
    \hline\hline
    $\rho$ & $0$ & UCT & $\mathbf{0}$ & $8$ & $10^5$ & $-$ & $-$ & $1$ & $53.9418$ &\bfseries 0.001752967\TBstrut\\
    \hline
    $\rho$ & $0.0025$ & UCT & $\mathbf{0.0025}$ & $8$ & $10^5$ & $-$ & $-$ & $1$ & $53.7572$ &\bfseries 0.00298362\TBstrut\\
    \hline
    $\rho$ & $100$   & UCT & $\mathbf{100}$ & $8$ & $10^5$ & $-$ & $-$ & $1$ & $49.0115$ & 0.600806429\TBstrut\\
    \hline\hline
    Core & $4$   & UCT & $\sqrt{2}$  & $\mathbf{4}$ & $10^5$ & $-$ & $-$ & $1$ & $49.0633$ & 0.355824737\TBstrut\\
    \hline
    Core & $1$   & UCT & $\sqrt{2}$  & $\mathbf{1}$ & $10^5$ & $-$ & $-$ & $1$ & $49.4586$ &\bfseries 0.000250096\TBstrut\\
    \hline\hline
    Tree Depth & $10^2$ & UCT & $\sqrt{2}$  & $8$ & $\mathbf{10^2}$ & $-$ & $-$ & $1$ & $45.3216$ &\bfseries 0.001406423 \TBstrut\\
    \hline
    Tree Depth & $10^3$ & UCT & $\sqrt{2}$  & $8$ & $\mathbf{10^3}$ & $-$ & $-$ & $1$ & $49.0667$ & 0.340845887 \TBstrut\\
    \hline
    Tree Depth & $10^7$ & UCT & $\sqrt{2}$  & $8$ & $\mathbf{10^7}$ & $-$ & $-$ & $1$ & $48.8048$ & 0.529105141 \TBstrut\\
    \hline\hline
    ConEx Timeout & $0.5$  & UCT  & $\sqrt{2}$  & $8$ & $10^5$ & $\mathbf{0.5}$ & $-$ & $1$ & $48.9121$ & 0.950464912 \TBstrut\\
    \hline
    ConEx Timeout & $1$  & UCT  & $\sqrt{2}$  & $8$ & $10^5$ & $\mathbf{1}$ & $-$ & $1$ & $49.0451$ & 0.422595538 \TBstrut\\
    \hline\hline
    SymEx Timeout & $5$  & UCT  & $\sqrt{2}$  & $8$ & $10^5$ & $-$ & $\mathbf{5}$ & $1$ & $49.0368$ & 0.3715003 \TBstrut\\
    \hline
    SymEx Timeout & $10$  & UCT  & $\sqrt{2}$  & $8$ & $10^5$ & $-$ & $\mathbf{10}$ & $1$ & $49.1712$ & 0.068867779 \TBstrut\\
    \hline\hline
    $\Nsamples$ & $3$  & UCT  & $\sqrt{2}$  & $8$ & $10^5$ & $-$ & $-$ & $\mathbf{3}$ & $49.0362$ & 0.569679957 \TBstrut\\
    \hline
    $\Nsamples$ & $5$  & UCT  & $\sqrt{2}$  & $8$ & $10^5$ & $-$ & $-$ & $\mathbf{5}$ & $49.1774$ & 0.31321928 \TBstrut\\
    \hline
\end{tabular}}

  \captionsetup{justification=centering}
  \caption{The sensitivity experiment: compared hyperparameter settings.
  Only one setting is changed in each alternative, indicated by the first column,
  including: the score function, exploration ratio ($\rho$), the number of cores (Core), 
  concrete execution timeout (ConEx Timeout), symbolic execution timeout (SymEx Timeout), 
  and the minimum number of samples per simulation ($\Nsamples$).
  The significant p-values ($<0.05$) are in bold font}
  \label{tbl: sensitivity}
\end{table*}

\section{Evaluation}\label{sec:evaluation}

We designed two kinds of experiments to evaluate \Legion's current design
and implementation. We sought to answer two fundamental questions,
respectively:
(1) is \Legion effective at generating high coverage test suites in fixed time,
as compared to other state-of-the-art tools? and (2) what is the effect of
the choices of \Legion's various hyperparameters on its performance, and
do there exist suitable default choices for these that are robust across
a variety of input programs?

{\bf Peer competition.} To answer the first question, \Legion competed in the \texttt{Cover-Branches} category of Test-Comp 2020 test generation competition.
\ifnew
\Legion has evolved since the competition; this paper reports the result of running the latest version of \Legion on the same bechmark suite \footnote{\url{https://github.com/sosy-lab/sv-benchmarks/tree/testcomp20}} on the same host machine with the same resources controlled by the same benchmarking framework as was used in Test-Comp 2020, thereby allowing our results to be compared against results obtained during the competition.
\fi

{\bf Sensitivity evaluation.} To answer the second question, we selected
a carefully chosen subset of the benchmark programs from the Test-Comp 2020
benchmarks, and then we evaluated the effect of varying \Legion's various
hyperparameters on its ability to perform on these programs.
\ifnew
Again, we report the results based on the most recent version of \Legion.
\else
\Legion's implementation has evolved since the competition and so we report results based on its most recent version.
\fi

\subsection{Experiment Setup} \label{sec: experiment setup}

To maximise reproducibility, we carried out both
evaluations using the \BenchExec framework \footnote{\url{https://gitlab.com/sosy-lab/software/benchexec/-/tree/2.5}} ~\cite{beyer2019BenchExec}.
It allows the usage of 8 cores and 15GB memory. Our evaluations used
Test-Comp 2020 benchmarks in which, for each, coverage
percentages computed by \texttt{testcov}\footnote{\url{https://gitlab.com/sosy-lab/software/test-suite-validator/-/tree/testcomp20}} were reported after 15 minutes.
There are $2531$ programs in total in the Test-Comp 2020 set,
sorted into $13$ \emph{suites}.
\Legion \footnote{\url{https://github.com/Alan32Liu/Legion/tree/TestComp2020-ASE2020v3}} requires two dependencies, \Angr \footnote{\url{https://github.com/Alan32Liu/angr/tree/TestComp2020-ASE2020}} and \Claripy \footnote{\url{https://github.com/Alan32Liu/claripy/tree/TestComp2020-ASE2020}}.
\iffalse % cut for space: Toby
The former is solely used to compute symbolic states along known concrete execution paths and invoke our implementation of \APPF in the latter.
\fi
% \Legion and our forks of both packages are publicly available at
% \ifnew
% \url{https://github.com/Alan32Liu/Legion/tree/TestComp2020-ASE2020v3}
% \else
% \url{https://github.com/Alan32Liu/Legion/tree/TestComp2020}
% \fi
% ,
% \url{https://github.com/Alan32Liu/angr/tree/TestComp2020-ASE2020}, and
% \url{https://github.com/Alan32Liu/claripy/tree/TestComp2020-ASE2020}, respectively.

\subsubsection{Peer competition} \label{sec: peer competition settings}

All tools to which we compared \Legion in the peer competition were seeded
with the same initial dummy input: 0. We purposefully
did not fine-tune the hyperparameters for the competition. 
\ifnew
For instance, we used the value of $\sqrt{2}$ for $\rho$, which is commonly used for UCT scores~\cite{MCTS-Survey}.
We limited the tree depth to be within $100000$, and the number of cores to be $8$, as they are sufficient for \Legion in most cases.
We did not time out symbolic or concrete execution, and run at least one input in each simulation stage ($\Nsamples=1$) like the original MCTS.
The results of our sensitivity experiments would later suggest that the performance of \Legion is not sensitive to most of these parameters when they were chosen reasonably.
(\cref{sec: sensitivity experiment results}).
\else
For instance,
we used the value of 
$\frac{1}{\sqrt{2}}$ for $\rho$, which is commonly
used for UCT scores~\cite{MCTS-Survey}.
The other hyperparameters were
chosen as: 10 seconds for both the symbolic and concrete execution timeouts,
$\Nsamples = 3$. The results of our sensitivity experiments
would later suggest that these values were far from optimal
(\cref{sec: sensitivity experiment results}).
\fi

Each experiment was conducted on the same host machine as the Test-Comp 2020 competition, with a 3.40 GHz Intel Xeon E3-1230 v5 CPU, running at 3800 MHz, with Turbo Boost disabled.

To support a fair and meaningful comparison, we excluded two benchmark suites, \texttt{BusyBox-MemSafety} and \texttt{SQLite-MemSafety}. All programs in the
former cause source code compilation errors due to name conflicts.
All tools scored close to $0$ on the latter suite, because of an unknown issue.

Since \Legion's goal is maximising coverage, our evaluation was restricted
to the coverage (as opposed to the error-finding) category (i.e. Cover-Branches) of Test-Comp 2020.

For the peer competition we ran \Legion in ``persistent'' mode (discussed in \cref{sec:optimisations}) to ensure that it would make full use of the time given to it for each benchmark program.

\subsubsection{Sensitivity Experiments} \label{sec: senstitivity experiement settings}

We evaluated \Legion's sensitivity to the various hyperparameters by
measuring its performance on the \texttt{Sequentialized} benchmark suite
from the Test-Comp 2020 benchmarks.
This benchmark suite was selected following the competition results, which
showed that \Legion had some room to improve its coverage score for this
suite, yet the competition result for it was not abysmal either. 
Hence,
effects on \Legion's performance here would be clearly visible in the final
coverage score obtained for this benchmark suite. This suite is also of
sufficient size, containing 81 programs that together comprise
26166 branches (323 branches on average per program).
\iffalse
, the top three in the competition respectively covered 62\%, 58\%, and 53\% with 73000, 73000, and 36000 CPU seconds. \Legion in the following experiments used 19000 CPU seconds}
\fi

\iffalse % no need to say this here since it is said later
The results indicated that \Legion is sensitive to one hyperparameter setting on this particular suite. But applying that setting on the whole competition in another experiment shows that it has very limited effect.
\fi

The various hyperparameter settings that we compared are listed in
\cref{tbl: sensitivity}.
For these experiments,
\Legion was run on a machine with a 2.50 GHz Intel Xeon Platinum 8180M CPU at 2494 MHz.
The ``persistent'' mode flag was not enabled during these sensitivity experiments to more accurately demonstrate how other hyperparameters affect the results.

% The host machine configured to be a 32 core Xeon VM with 200 GB memory. But the resource of each experiment is controlled to be 

% Namely, we choose the following benchmark suites from the competition:

% \begin{itemize}
%   \item \texttt{Arrays}. We got the third place of this suite in the competition, very close to the first two and way ahead of the fourth. We choose it because \angr encounters the early concretisation in many benchmarks of this suite.
%   \item \texttt{DeviceDriversLinux}. We choose this benchmark because we won the first place of this suite in the competition, which means either we got lucky for happen to choose a suitable set of hyperparameter settings, or \Legion's algorithm is well designed as a fuzzing framework.
%   \item \texttt{Sequentialised}. We choose it because the competition result shows \Legion has room for improvement on this suite. We want to investigate how much can it improve with different parameters.
%   \item {\color{red} Others with similar reasons?}
% \end{itemize}

\subsection{Results} \label{sec: experiment results}

\subsubsection{Peer Competition} \label{sec: peer competition results}

The normalised\footnote{https://test-comp.sosy-lab.org/2020/rules.php\#meta} results from the peer competition appear in
\cref{tbl: competition result}.  Each benchmark suite is listed in the first
column, with the number of programs it contains in brackets.
That number is also a theoretical upper bound on the total coverage score
that can be achieved for each category. The following columns list the
score of each competitor on each suite, where the score is the average of
the coverage score (between 0 and 1) achieved on each program in the suite.

\begin{table*}[!ht]
  \centering
  \resizebox{1.01\linewidth}{!}{%
  
\begin{tabular}{|r|lllllllll|}
    %   \hline
    %   \multicolumn{10}{|c|}{Accumulated coverage score for each benchmark suite(The higher the better)} \\
    \hline
    Participants & \Legion* & \texttt{KLEE}~\cite{cadar2008klee} & \texttt{CoVeriTest}~\cite{beyer2019coveritest} & \texttt{HybridTiger}~\cite{ruland2020hybridtiger} & \texttt{LibKluzzer}~\cite{le2020llvm} & \texttt{PRTest} & \texttt{Symbiotic}~\cite{Symbiotic7} & \texttt{Tracer-X}~\cite{jaffar2020tracerx}  & \texttt{VeriFuzz}~\cite{VeriFuzz} \\
    \hline\hline
    Arrays ($243$) & $\mathbf{123}$ & $121$ & $123$ & $126$ & $201$ & $93.4$ & $75.4$ & $116$ & $200$ \\
    \hline
    Floats ($212$) & $\mathbf{105}$ & $59.7$ & $107$ & $107$ & $107$ & $51.2$ & $56.4$ & $50.8$ & $97.7$ \\
    \hline
    Heap ($139$) & $\mathbf{102}$ & $101$ & $91.6$ & $98.4$ & $104$ & $51.6$ & $82.7$ & $94.7$ & $103$ \\
    \hline
    Loops ($123$) & $\mathbf{97.6}$ & $93$ & $95.8$ & $92.5$ & $101$ & $54.1$ & $71.3$ & $87.3$ & $97.1$ \\
    \hline
    Recursive ($38$) & $\mathbf{34.3}$ & $32.1$ & $30.9$ & $30.8$ & $34.3$ & $9.85$ & $21.3$ & $27.6$ & $34.2$ \\
    \hline
    DeviceDriversLinux64 ($290$) & $\mathbf{22.9}$ & $21.7$ & $23.2$ & $22.1$ & $22.1$ & $9.24$ & $21$ & $22$ & $22.7$ \\
    \hline
    MainHeap ($231$) & $\mathbf{161}$ & $153$ & $181$ & $180$ & $195$ & $53.3$ & $129$ & $175$ & $195$ \\
    \hline
    BitVectors ($36$) & $23.2$ & $25.7$ & $27.7$ & $27$ & $28.2$ & $3.43$ & $18.6$ & $27.6$ & $28.1$ \\
    \hline
    ControlFlow ($19$) & $6.71$ & $12.6$ & $13.9$ & $13.9$ & $13.8$ & $0.558$ & $1.68$ & $12.7$ & $13.4$ \\
    \hline
    ECA ($1046$) & $729$ & $957$ & $811$ & $766$ & $962$ & $489$ & $639$ & $932$ & $963$ \\
    \hline
    Sequentialized ($81$) & $49.7$ & $52.5$ & $57.4$ & $43.5$ & $58.2$ & $5.28$ & $14.5$ & $15.7$ & $62.5$ \\
    \hline
    Normailised Average** & $1455.908002$ & $1515.97673$ & $1588.862835$ & $1541.850942$ & $1759.050529$ & $584.57209$ & $969.1994602$ & $1382.674571$ & $1746.261043$ \\
    \hline
    % {\color{red} Normailised Average***} & $1502.984227$ & $1492.940073$ & $1555.807154$ & $1504.493807$ & $1742.850222$ & $630.8958136$ & $1036.312556$ & $1346.157462$ & $1734.025271$ \\
    % \hline
    % {\color{red} Normailised Average****} & $1452.564362$ & $1443.719195$ & $1504.030926$ & $1454.700936$ & $1688.051552$ & $610.038214$ & $997.2578185$ & $1299.7722$ & $1678.917913$ \\
    % \hline
\end{tabular}

  }
  \caption{Normalised accumulated coverage scores for each benchmark suite (The higher the better).
  \Legion outperformed \texttt{KLEE} in $7$ out of $11$ benchmark suites (in bold font). \small * The results are reproduced with the latest version of \Legion; 
  \small ** Excluded two benchmark suites to avoid noise: \texttt{BusyBox-MemSafety}, \texttt{SQLite-MemSafety}
  % *** Excluded three benchmark suites: \texttt{BusyBox-MemSafety}, \texttt{SQLite-MemSafety}, \texttt{ControlFlow}\\
  % **** Excluded four benchmark suites: \texttt{BusyBox-MemSafety}, \texttt{SQLite-MemSafety}, \texttt{ControlFlow}, \texttt{DeviceDriversLinux64-ReachSafety}
  }
  % \small\textsuperscript{a=} https://test-comp.sosy-lab.org/2020/rules.php#meta
  \label{tbl: competition result}
\end{table*}

% \begin{table*}[!ht]
%   % \setlength\tabcolsep{0.75pt}
%   \centering
%   \resizebox{1.01\linewidth}{!}{%
%   \input{Tables/CompetitionResult2.tex}
%   }
%   \caption{Normalised accumulated coverage scores for each benchmark suite (The higher the better).\\
%   {\color{red}
%   \small* The results are reproduced with the latest version of \Legion \\
%   % ** Excluded two benchmark suites to avoid noise: \texttt{BusyBox-MemSafety}, \texttt{SQLite-MemSafety}, \texttt{}
%   \small*** Excluded three benchmark suites: \texttt{BusyBox-MemSafety}, \texttt{SQLite-MemSafety}, \texttt{ControlFlow}\\
%   % **** Excluded four benchmark suites: \texttt{BusyBox-MemSafety}, \texttt{SQLite-MemSafety}, \texttt{ControlFlow}, \texttt{DeviceDriversLinux64-ReachSafety}
%   }
%   }
%   \label{tbl: competition result 2}
% \end{table*}

The peer experiment demonstrates that \Legion is effective across a wide
variety of programs, despite the simplicity of its current heuristics
and limitations of its implementation.
\Legion is competitive in many suites of Test-Comp~2020, achieving within~90\% of the best score in~5 of the~11 suites in the \texttt{Cover-Branches} category.
\Legion outperformed all other tools on some benchmarks (discussed below), which highlights the general effectiveness of the approach.

\subsubsection{Sensitivity Experiments} \label{sec: sensitivity experiment results}

For each of the hyperparameter settings investigated,
\cref{tbl: sensitivity} reports the 
\texttt{Sequentialized} suite
coverage result. It also reports the p-value obtained
by applying Student's t-test (paired, two-tailed) to determine whether the
difference from the baseline was statistically significant.
Conventionally, we claim the difference is significant when that value
is less than 0.05. 

% \begin{table*}[!ht]
%   \input{Tables/SensitivityResult.tex}
%   \caption{Accumulated coverage scores and and t-Test p-value of benchmark suite \texttt{Sequentialized}}
%   \label{tbl: sensitivityResult}
% \end{table*}
As to be expected, These results indicate UCT scoring superior to the baseline random selection.
\Legion's performance is also sensitive to the choice of its exploration ratio $\rho$.
Small values of $\rho$ near 0.0025 appear to work best, based on these results, and outperform pure
exploitation ($\rho = 0$).
For currently unknown reasons, using 1 core appears to have a slightly better performance.
The choices of other hyperparameters, including the tree depth, concrete execution timeout, symbolic execution timeout, and the minimum number of inputs to generate for each \APPF ($\Nsamples$), appear to be not influential when chosen within a reasonable range.
% (discussed in \cref{sec: discussion})

However, the value of $\rho$ still exhibited very limited effect across all suites when we were measuring \Legion's overall performance to rule out overfitting. The most preferable setting above ($\rho=0.0025$) improved the overall competition score by merely $1.04\%$.

% \cref{tbl: AlternativesensitivityResult} reports the best choice of
% hyperparameter settings from \cref{tbl: sensitivityResult} obtained for
% the \texttt{Sequentialized} benchmark suite
% also leads to better results for the \texttt{Loops} suite and the \texttt{Recursive} suite.

% \begin{table*}[ht]
%   \qquad \qquad \begin{minipage}[b]{0.35\textwidth}
%     \centering
%     \input{Tables/AlternativeSensitivityResult.tex}
%     % \vspace*{-.3cm}
%   \end{minipage}
%   \hfill
%   \begin{minipage}[b]{0.5\textwidth}
%     \centering
%     \input{Tables/ThirdSensitivityResult.tex}
%    \end{minipage}
%   % \vspace*{-.5cm}
%   \caption{Accumulated coverage scores and and t-Test p-value of benchmark suite \texttt{Loops} (left) and \texttt{Recursive} (right)}
%   \label{tbl: AlternativesensitivityResult}
% \end{table*}

\subsection{Discussion}
\label{sec: discussion}

\Legion inherits limitations from \Angr and MCTS.

\Angr, as symbolic execution backend of \Legion, eagerly concretises size values used for dynamic memory allocations, 
which then causes it to conclude erroneously that certain observed concrete execution paths are infeasible.
The design of \Legion's MCTS allows it to work around this limitation:
\Legion first detects this case from the mismatch between symbolic and concrete execution, then 
omits the erroneous programs states from the selection stage. When MCTS predicts new paths under them, \Legion invokes \APPF on their parents instead. With this mitigation, \Legion achieved full coverage on \texttt{loops/sum\_array-1.c} (in contrast to all other tools) despite the dynamic allocations.

\Angr uses VEX IR
%\footnote{\url{https://docs.angr.io/advanced-topics/ir}}
which prior work~\cite{yun2018qsym}
has noted leads to increased overheads in symbolic execution. \Legion mitigates this by performing symbolic execution lazily, only computing the symbolic states of nodes selected by MCTS.

\Angr cannot model arrays that have more than $4096$ ($0x1000$) elements.
In particular, it failed on $16$ benchmarks in suite \texttt{Arrays} that involves multidimensional arrays.

\Angr only supports 22 out of the 337 system calls in the Linux kernel 2.6~\cite{yun2018qsym,DigFuzz}, and is known to scale poorly to large
programs~\cite{TFUZZ}.

% Another problem is the allocation of such dynamic memory when the size happens to be randomly chosen to be a very large value (in particular in 64 Bit mode).
% This may cause out-of-memory in the concrete executions, or alternatively lead to hugely long execution traces.
% The mitigation outlined above helps here, too.

The nature of MCTS makes \Legion less suitable for exploring spaces where the reward is rare.
For example, benchmarks in suite \texttt{ControlFlow} are full of long sequences of equality constraints that are satisfied by few inputs, making new path discovery via sampling (i.e. \APPF) very rare.
This benchmark suite happens to be an extreme negative example of the trade-off in depth (discussed in \cref{sec: Overview}) that \Legion intends to balance where fuzzing the parent gives no reward.
Expensive constraint solving also adds extra cost to sampling. 
In particular, \Claripy spent~$14.82$ seconds on average to solve each constraint in this suite.
Generally, invoking \APPF at intermediate nodes with less complicated constraints to generate mutations can amortise this cost.
However, although on average \Legion generated $17.37$ mutations per program at negligible costs ($10^{-5}$ seconds per mutation) and $99.39\%$ of them preserved their paths, none of them found any new path due to the equality constraints.
To a certain extent, this could possibly be mitigated by decreasing \Legion's exploration ratio (constant $\rho$ in the UCT score) but such fine-tuning would overfit \Legion to the competition benchmark suites.

To analyse the impact of this extreme example, we re-normalised the scores of the peer competition 
% in \cref{tbl: competition result 2} 
with this suite excluded. 
It shows that the overall normalised score of \Legion ($1502.98$) is higher than \texttt{KLEE} ($1492.94$) and close to \texttt{HybridTiger} ($1504.49$).

MCTS works much better on less extreme examples. Benchmarks \texttt{seq-mthreaded/pals\_lcr-var-start-time*.c} interleave equality constraints with range constraints, making new path discovery less rare ($11.77\%$ of the new paths are found via sampling at a path-preserving rate of $93.49\%$) and less expensive (each constraint solving took $0.81$ seconds on average). As a result, \Legion performed roughly the same as \texttt{CoveriTest} on these $12$ benchmark programs and largely outperformed all other competitors. Similarly, \Legion fully covered the sample program \texttt{Ackermann02.c} (shown in \cref{lst:ackermann02}) with $77.63\%$ of the new paths uncovered by mutations from \APPF ($71.88\%$ of them preserving their path prefix) at an average cost of $1.10*10^{-5}$ seconds, approximately $1000$ times faster than constraint solving via \Claripy.

\section{Related Work} \label{sec: Related Work}

We consider recent work that adopts similar ideas to \Legion, at the
intersection of fuzzing and concolic execution. 

Like \Legion, Driller's~\cite{Driller} design is also inspired by wanting to
harness the complementary strengths of concolic execution and fuzzing.
It augments AFL~\cite{AFL} by detecting when fuzzing gets stuck and then
applying concolic execution (constraint solving) to generate inputs to feed
back to the fuzzer, to allow it to traverse choke points.
Unlike our \APPF, however, little attempt is made to ensure that subsequent
mutations made by the fuzzer to each solver-generated input preserves the
input's paths condition. Thus AFL might undo some of the hard work done by
the solver.

\Legion uses rewards from past \APPF to predict where to perform concolic
execution, modelling automated test generation as Monte Carlo
Tree Search.
The authors
in~\cite{MDPC} formally model programs as Markov Decision Processes with
Costs. Their model is used to decide when to perform input generation via
constraint solving vs.\ when to generate inputs via pure random fuzzing.
Like \Legion, their approach tracks the
past rewards from concrete execution to decide when to
perform fuzzing. However, unlike \Legion their fuzzing approach is purely
random.
\Legion applies the UCT algorithm to make choices about which parts of the
program to target at each iteration, while the approach of~\cite{MDPC}
instead applies a greedy algorithm to estimate the rewards of random input
generation. 
The authors' approach~\cite{MDPC}, unlike \Legion, applies a static heuristic
to estimate the cost of
constraint solving in order to make an informed decision between the
precision and expense of constraint solving vs.\ the imprecision and
efficiency of pure random fuzzing. \Legion eschews this all-or-nothing choice
altogether, by blending fuzzing and concolic execution via
\APPF to avoid the drawbacks of both while retaining their strengths.
\iffalse % cut for space
An interesting direction for future work would be to incorporate a cost
model into \Legion's algorithm, to predict when \APPF is likely to be
expensive.
\fi

DigFuzz~\cite{DigFuzz} extends Driller and,
similar to \Legion, collects statistics to decide which program states
would benefit most from concolic execution, via Monte Carlo simulations.

While \Legion collects coverage statistics, DigFuzz instead focuses on learning
which parts of the program are most difficult for the fuzzer to traverse,
and only then does it decide
to invoke the solver, on the assumption that solving is expensive and should
only be used when necessary.
Whereas \Legion estimates the likelihood of reward (uncovering new execution paths via directed \APPF)
for each node, based on past rewards from \APPF, DigFuzz instead
estimates the probability that the undirected fuzzer will be able to traverse each path,
based on the fuzzer's past performance. It collects statistics about how
often each branch is traversed by the fuzzer. Then the probability of
traversing a particular path is computed by multiplying together estimates
of the probabilities on each of the branches along that path.

\iffalse % Toby: cut this since we don't justify our assumption much either
Multiplying probabilities together like this is sound only when the corresponding events
are independent. It is unclear whether that assumption is true in reality.
when using AFL.
On the other hand, \Legion's MCTS makes no similar assumption. Instead, MCTS
assumes that simulations are i.i.d. While \Legion does not guarantee that
this is the case, our \APPF implementation was designed to try to meet it
in practice.
\fi

So while both methods use a form of Monte Carlo simulation to track statistics
to decide how to generate inputs for the program,
they do so for quite different purposes
and using methods that appear relatively complementary.  This comes
from their different design philosophies: DigFuzz wants to minimise solver
use as much as possible, and is happy to spend more time on repeated traversal
of the paths by the cheap fuzzer.
\Legion instead is willing to use the solver more
often (although not as often as in traditional concolic testing), to avoid
concrete executions that often repeat the same execution paths.

MCTS has also been used recently to guide traditional concolic
execution~\cite{MCTSSE}. The authors use MCTS to pick which branch to
concolically execute. Unlike \Legion, this requires symbolic execution along
the entire path, which \Legion can avoid since it only ever performs
symbolic execution lazily (\cref{sec: tree nodes}). It also requires one
solver call for each concrete execution. \Legion avoids this by using
\APPF instead, which
 generates increasingly more inputs for each solver invocation, at the price
 of not guaranteeing that all generated inputs will satisfy the path
 constraint. Finally, while \Legion's score function applies UCT to
estimate rewards in terms of discovering new execution paths,
it is not clear what is the score function used in~\cite{MCTSSE}.

Finally, we note that other work has also attempted to improve
fuzzing by adopting methods from machine learning. For instance,
Learn\&Fuzz~\cite{godefroid2017learn} trains a neural network to represent
the program under test, to aid fuzzing. The Angora~\cite{chen2018angora}
fuzzer applies gradient descent as an input mutation strategy to help it
traverse conditional branches, while AFLGo~\cite{bohme2017AFLGo} uses simulated
annealing to guide directed fuzzing. In contrast, \Legion adopts MCTS with
\APPF as a principled unification of fuzzing and concolic execution.

% % \textbf{$\bullet$ Summary of main contribution}
% \textbf{Contributions}: The contribution of this paper are as follows:
% \begin{itemize}
%     \item First, it presents a novel approach to integrate symbolic execution, fuzzing and machine learning
%     \item Second, it easily adapts to different implementations of symbolic execution and path preserving fuzzing as modules for performance improvement or comparison.
%     \item Third, its experimental performance demonstrates its effectiveness and efficiency in generating high code coverage input suites.
% \end{itemize}

\iffalse % cut for space
\input{FutureWork}
\fi

\section{Conclusion} \label{sec: conclusion}

How can we maximise code coverage by generalising concolic execuiton and fuzzing? 
\Legion shows a possible solution with its variation of the Monte Carlo Tree Search algorithm. By tracking past rewards from \APPF,
\Legion learns where
concolic execution effort should be spent, namely on those parts of the
program in which new execution paths have been observed in the past.
Symbolic execution then returns the favour by computing new promising
symbolic states to support accurate \APPF.
Which states to prefer is estimated using the UCT algorithm,
which navigates the exploration versus exploitation trade-off.

% We presented \Legion, which utilises Monte Carlo tree search and introduces
% approximate path-preserving fuzzing (\APPF) to concolic execution and coverage-guided
% fuzzing. \Legion's MCTS means it adopts a best-first search strategy to
% concolic testing, while its \APPF ensures it can generate inputs with higher
% efficiency than traditional concolic testing.

We demonstrated a simple reward heuristic to maximise discovery of new
execution paths, and showed how this could be encoded in MCTS's traditional
UCT score function. We evaluated \Legion, both to measure its performance
against its peers and to understand its sensitivity to hyperparameters.
We found that \Legion was effective on a wide variety
of benchmark programs without being overly sensitive to
  reasonable choices for its hyperparameters.

\Legion provides a new framework for investigating trade-offs between
traditional testing approaches, and the incorporation of further statistical
learning methods to assist automated test generation. Our results demonstrate
the practicality and the promise of the ideas it embodies.

\iffalse
\textit{\color{blue} At most 10 pages for main text, 2 pages for references}
\textit{\color{blue} Use HotCRP format checker}
\fi

%\clearpage

\bibliographystyle{ACM-Reference-Format}

\bibliography{Legion-ASE}


\begin{thebibliography}{37}

%%% the Web address.

\ifx \showCODEN    \undefined \def \showCODEN     #1{\unskip}     \fi
\ifx \showDOI      \undefined \def \showDOI       #1{#1}\fi
\ifx \showISBNx    \undefined \def \showISBNx     #1{\unskip}     \fi
\ifx \showISBNxiii \undefined \def \showISBNxiii  #1{\unskip}     \fi
\ifx \showISSN     \undefined \def \showISSN      #1{\unskip}     \fi
\ifx \showLCCN     \undefined \def \showLCCN      #1{\unskip}     \fi
\ifx \shownote     \undefined \def \shownote      #1{#1}          \fi
\ifx \showarticletitle \undefined \def \showarticletitle #1{#1}   \fi
\ifx \showURL      \undefined \def \showURL       {\relax}        \fi
% invisible to TeX
\providecommand\bibfield[2]{#2}
\providecommand\bibinfo[2]{#2}
\providecommand\natexlab[1]{#1}
\providecommand\showeprint[2][]{arXiv:#2}

\bibitem[\protect\citeauthoryear{Basak~Chowdhury, Medicherla, and
  R}{Basak~Chowdhury et~al\mbox{.}}{2019}]%
        {VeriFuzz}
\bibfield{author}{\bibinfo{person}{Animesh Basak~Chowdhury},
  \bibinfo{person}{Raveendra~Kumar Medicherla}, {and}
  \bibinfo{person}{Venkatesh R}.} \bibinfo{year}{2019}\natexlab{}.
\newblock \showarticletitle{VeriFuzz: Program Aware Fuzzing}. In
  \bibinfo{booktitle}{\emph{Tools and Algorithms for the Construction and
  Analysis of Systems}}, \bibfield{editor}{\bibinfo{person}{Dirk Beyer},
  \bibinfo{person}{Marieke Huisman}, \bibinfo{person}{Fabrice Kordon}, {and}
  \bibinfo{person}{Bernhard Steffen}} (Eds.). \bibinfo{publisher}{Springer
  International Publishing}, \bibinfo{address}{Cham},
  \bibinfo{pages}{244--249}.
\newblock
\showISBNx{978-3-030-17502-3}


\bibitem[\protect\citeauthoryear{Beyer}{Beyer}{2020}]%
        {TESTCOMP20}
\bibfield{author}{\bibinfo{person}{D. Beyer}.} \bibinfo{year}{2020}\natexlab{}.
\newblock \showarticletitle{Second Competition on Software Testing: Test-Comp
  2020}. In \bibinfo{booktitle}{\emph{Proc.\ FASE}}
  \emph{(\bibinfo{series}{LNCS~})}. \bibinfo{publisher}{Springer}.
\newblock
\urldef\tempurl%
\url{https://www.sosy-lab.org/research/pub/2020-FASE.Second_Competition_on_Software_Testing_Test-Comp_2020.pdf}
\showURL{%
\tempurl}


\bibitem[\protect\citeauthoryear{Beyer and Jakobs}{Beyer and Jakobs}{2019}]%
        {beyer2019coveritest}
\bibfield{author}{\bibinfo{person}{Dirk Beyer} {and}
  \bibinfo{person}{Marie-Christine Jakobs}.} \bibinfo{year}{2019}\natexlab{}.
\newblock \showarticletitle{CoVeriTest: Cooperative verifier-based testing}. In
  \bibinfo{booktitle}{\emph{International Conference on Fundamental Approaches
  to Software Engineering}}. Springer, \bibinfo{pages}{389--408}.
\newblock


\bibitem[\protect\citeauthoryear{Beyer, L{\"o}we, and Wendler}{Beyer
  et~al\mbox{.}}{2019}]%
        {beyer2019BenchExec}
\bibfield{author}{\bibinfo{person}{Dirk Beyer}, \bibinfo{person}{Stefan
  L{\"o}we}, {and} \bibinfo{person}{Philipp Wendler}.}
  \bibinfo{year}{2019}\natexlab{}.
\newblock \showarticletitle{Reliable benchmarking: Requirements and solutions}.
\newblock \bibinfo{journal}{\emph{International Journal on Software Tools for
  Technology Transfer}} \bibinfo{volume}{21}, \bibinfo{number}{1}
  (\bibinfo{year}{2019}), \bibinfo{pages}{1--29}.
\newblock


\bibitem[\protect\citeauthoryear{B{\"o}hme, Pham, Nguyen, and
  Roychoudhury}{B{\"o}hme et~al\mbox{.}}{2017b}]%
        {bohme2017AFLGo}
\bibfield{author}{\bibinfo{person}{Marcel B{\"o}hme},
  \bibinfo{person}{Van-Thuan Pham}, \bibinfo{person}{Manh-Dung Nguyen}, {and}
  \bibinfo{person}{Abhik Roychoudhury}.} \bibinfo{year}{2017}\natexlab{b}.
\newblock \showarticletitle{Directed greybox fuzzing}. In
  \bibinfo{booktitle}{\emph{Proceedings of the 2017 ACM SIGSAC Conference on
  Computer and Communications Security}}. \bibinfo{pages}{2329--2344}.
\newblock


\bibitem[\protect\citeauthoryear{B{\"o}hme, Pham, and Roychoudhury}{B{\"o}hme
  et~al\mbox{.}}{2017a}]%
        {bohme2017AFLFast}
\bibfield{author}{\bibinfo{person}{Marcel B{\"o}hme},
  \bibinfo{person}{Van-Thuan Pham}, {and} \bibinfo{person}{Abhik
  Roychoudhury}.} \bibinfo{year}{2017}\natexlab{a}.
\newblock \showarticletitle{Coverage-based greybox fuzzing as {M}arkov chain}.
\newblock \bibinfo{journal}{\emph{IEEE Transactions on Software Engineering}}
  \bibinfo{volume}{45}, \bibinfo{number}{5} (\bibinfo{year}{2017}),
  \bibinfo{pages}{489--506}.
\newblock


\bibitem[\protect\citeauthoryear{{Browne}, {Powley}, {Whitehouse}, {Lucas},
  {Cowling}, {Rohlfshagen}, {Tavener}, {Perez}, {Samothrakis}, and
  {Colton}}{{Browne} et~al\mbox{.}}{2012}]%
        {MCTS-Survey}
\bibfield{author}{\bibinfo{person}{C.~B. {Browne}}, \bibinfo{person}{E.
  {Powley}}, \bibinfo{person}{D. {Whitehouse}}, \bibinfo{person}{S.~M.
  {Lucas}}, \bibinfo{person}{P.~I. {Cowling}}, \bibinfo{person}{P.
  {Rohlfshagen}}, \bibinfo{person}{S. {Tavener}}, \bibinfo{person}{D. {Perez}},
  \bibinfo{person}{S. {Samothrakis}}, {and} \bibinfo{person}{S. {Colton}}.}
  \bibinfo{year}{2012}\natexlab{}.
\newblock \showarticletitle{A Survey of Monte Carlo Tree Search Methods}.
\newblock \bibinfo{journal}{\emph{IEEE Transactions on Computational
  Intelligence and AI in Games}} \bibinfo{volume}{4}, \bibinfo{number}{1}
  (\bibinfo{date}{March} \bibinfo{year}{2012}), \bibinfo{pages}{1--43}.
\newblock
\showISSN{1943-068X}
\urldef\tempurl%
\url{https://doi.org/10.1109/TCIAIG.2012.2186810}
\showDOI{\tempurl}


\bibitem[\protect\citeauthoryear{Cadar, Dunbar, Engler, et~al\mbox{.}}{Cadar
  et~al\mbox{.}}{2008}]%
        {cadar2008klee}
\bibfield{author}{\bibinfo{person}{Cristian Cadar}, \bibinfo{person}{Daniel
  Dunbar}, \bibinfo{person}{Dawson~R Engler}, {et~al\mbox{.}}}
  \bibinfo{year}{2008}\natexlab{}.
\newblock \showarticletitle{KLEE: Unassisted and Automatic Generation of
  High-Coverage Tests for Complex Systems Programs.}. In
  \bibinfo{booktitle}{\emph{OSDI}}, Vol.~\bibinfo{volume}{8}.
  \bibinfo{pages}{209--224}.
\newblock


\bibitem[\protect\citeauthoryear{Chalupa, Ja{\v{s}}ek, Tomovi{\v{c}},
  Hru{\v{s}}ka, {\v{S}}okov{\'a}, Ayaziov{\'a}, Strej{\v{c}}ek, and
  Vojnar}{Chalupa et~al\mbox{.}}{2020}]%
        {Symbiotic7}
\bibfield{author}{\bibinfo{person}{Marek Chalupa},
  \bibinfo{person}{Tom{\'a}{\v{s}} Ja{\v{s}}ek},
  \bibinfo{person}{Luk{\'a}{\v{s}} Tomovi{\v{c}}}, \bibinfo{person}{Martin
  Hru{\v{s}}ka}, \bibinfo{person}{Veronika {\v{S}}okov{\'a}},
  \bibinfo{person}{Paul{\'i}na Ayaziov{\'a}}, \bibinfo{person}{Jan
  Strej{\v{c}}ek}, {and} \bibinfo{person}{Tom{\'a}{\v{s}} Vojnar}.}
  \bibinfo{year}{2020}\natexlab{}.
\newblock \showarticletitle{Symbiotic 7: Integration of Predator and More}. In
  \bibinfo{booktitle}{\emph{Tools and Algorithms for the Construction and
  Analysis of Systems}}, \bibfield{editor}{\bibinfo{person}{Armin Biere} {and}
  \bibinfo{person}{David Parker}} (Eds.). \bibinfo{publisher}{Springer
  International Publishing}, \bibinfo{address}{Cham},
  \bibinfo{pages}{413--417}.
\newblock
\showISBNx{978-3-030-45237-7}


\bibitem[\protect\citeauthoryear{Chen and Chen}{Chen and Chen}{2018}]%
        {chen2018angora}
\bibfield{author}{\bibinfo{person}{Peng Chen} {and} \bibinfo{person}{Hao
  Chen}.} \bibinfo{year}{2018}\natexlab{}.
\newblock \showarticletitle{Angora: Efficient fuzzing by principled search}. In
  \bibinfo{booktitle}{\emph{2018 IEEE Symposium on Security and Privacy (SP)}}.
  IEEE, \bibinfo{pages}{711--725}.
\newblock


\bibitem[\protect\citeauthoryear{{Dutra}, {Laeufer}, {Bachrach}, and
  {Sen}}{{Dutra} et~al\mbox{.}}{2018}]%
        {QuickSampler}
\bibfield{author}{\bibinfo{person}{R. {Dutra}}, \bibinfo{person}{K. {Laeufer}},
  \bibinfo{person}{J. {Bachrach}}, {and} \bibinfo{person}{K. {Sen}}.}
  \bibinfo{year}{2018}\natexlab{}.
\newblock \showarticletitle{Efficient Sampling of SAT Solutions for Testing}.
  In \bibinfo{booktitle}{\emph{2018 IEEE/ACM 40th International Conference on
  Software Engineering (ICSE)}}. \bibinfo{pages}{549--559}.
\newblock
\showISSN{1558-1225}
\urldef\tempurl%
\url{https://doi.org/10.1145/3180155.3180248}
\showDOI{\tempurl}


\bibitem[\protect\citeauthoryear{Gan, Zhang, Qin, Tu, Li, Pei, and Chen}{Gan
  et~al\mbox{.}}{2018}]%
        {gan2018collafl}
\bibfield{author}{\bibinfo{person}{Shuitao Gan}, \bibinfo{person}{Chao Zhang},
  \bibinfo{person}{Xiaojun Qin}, \bibinfo{person}{Xuwen Tu},
  \bibinfo{person}{Kang Li}, \bibinfo{person}{Zhongyu Pei}, {and}
  \bibinfo{person}{Zuoning Chen}.} \bibinfo{year}{2018}\natexlab{}.
\newblock \showarticletitle{Collafl: Path sensitive fuzzing}. In
  \bibinfo{booktitle}{\emph{2018 IEEE Symposium on Security and Privacy (SP)}}.
  IEEE, \bibinfo{pages}{679--696}.
\newblock


\bibitem[\protect\citeauthoryear{Godefroid, Klarlund, and Sen}{Godefroid
  et~al\mbox{.}}{2005}]%
        {DART}
\bibfield{author}{\bibinfo{person}{Patrice Godefroid}, \bibinfo{person}{Nils
  Klarlund}, {and} \bibinfo{person}{Koushik Sen}.}
  \bibinfo{year}{2005}\natexlab{}.
\newblock \showarticletitle{DART: Directed Automated Random Testing}. In
  \bibinfo{booktitle}{\emph{Proceedings of the 2005 ACM SIGPLAN Conference on
  Programming Language Design and Implementation}} (Chicago, IL, USA)
  \emph{(\bibinfo{series}{PLDI ’05})}. \bibinfo{publisher}{Association for
  Computing Machinery}, \bibinfo{address}{New York, NY, USA},
  \bibinfo{pages}{213–223}.
\newblock
\showISBNx{1595930566}
\urldef\tempurl%
\url{https://doi.org/10.1145/1065010.1065036}
\showDOI{\tempurl}


\bibitem[\protect\citeauthoryear{Godefroid, Levin, and Molnar}{Godefroid
  et~al\mbox{.}}{2012}]%
        {SAGE}
\bibfield{author}{\bibinfo{person}{Patrice Godefroid},
  \bibinfo{person}{Michael~Y. Levin}, {and} \bibinfo{person}{David Molnar}.}
  \bibinfo{year}{2012}\natexlab{}.
\newblock \showarticletitle{SAGE: Whitebox Fuzzing for Security Testing}.
\newblock \bibinfo{journal}{\emph{Queue}} \bibinfo{volume}{10},
  \bibinfo{number}{1} (\bibinfo{date}{Jan.} \bibinfo{year}{2012}),
  \bibinfo{pages}{20–27}.
\newblock
\showISSN{1542-7730}
\urldef\tempurl%
\url{https://doi.org/10.1145/2090147.2094081}
\showDOI{\tempurl}


\bibitem[\protect\citeauthoryear{Godefroid, Peleg, and Singh}{Godefroid
  et~al\mbox{.}}{2017}]%
        {godefroid2017learn}
\bibfield{author}{\bibinfo{person}{Patrice Godefroid}, \bibinfo{person}{Hila
  Peleg}, {and} \bibinfo{person}{Rishabh Singh}.}
  \bibinfo{year}{2017}\natexlab{}.
\newblock \showarticletitle{Learn\&fuzz: Machine learning for input fuzzing}.
  In \bibinfo{booktitle}{\emph{2017 32nd IEEE/ACM International Conference on
  Automated Software Engineering (ASE)}}. IEEE, \bibinfo{pages}{50--59}.
\newblock


\bibitem[\protect\citeauthoryear{Jaffar, Maghareh, Godboley, and Ha}{Jaffar
  et~al\mbox{.}}{2020}]%
        {jaffar2020tracerx}
\bibfield{author}{\bibinfo{person}{Joxan Jaffar}, \bibinfo{person}{Rasool
  Maghareh}, \bibinfo{person}{Sangharatna Godboley}, {and}
  \bibinfo{person}{Xuan-Linh Ha}.} \bibinfo{year}{2020}\natexlab{}.
\newblock \showarticletitle{TracerX: Dynamic symbolic execution with
  interpolation (competition contribution)}. In
  \bibinfo{booktitle}{\emph{International Conference on Fundamental Approaches
  to Software Engineering}}. Springer, \bibinfo{pages}{530--534}.
\newblock


\bibitem[\protect\citeauthoryear{Kocsis and Szepesv{\'a}ri}{Kocsis and
  Szepesv{\'a}ri}{2006}]%
        {kocsis2006bandit}
\bibfield{author}{\bibinfo{person}{Levente Kocsis} {and} \bibinfo{person}{Csaba
  Szepesv{\'a}ri}.} \bibinfo{year}{2006}\natexlab{}.
\newblock \showarticletitle{Bandit based {M}onte-{C}arlo planning}. In
  \bibinfo{booktitle}{\emph{European Conference on Machine Learning}}.
  Springer, \bibinfo{pages}{282--293}.
\newblock


\bibitem[\protect\citeauthoryear{Le}{Le}{2020}]%
        {le2020llvm}
\bibfield{author}{\bibinfo{person}{Hoang~M Le}.}
  \bibinfo{year}{2020}\natexlab{}.
\newblock \showarticletitle{Llvm-based hybrid fuzzing with LibKluzzer
  (competition contribution)}. In \bibinfo{booktitle}{\emph{International
  Conference on Fundamental Approaches to Software Engineering}}. Springer,
  \bibinfo{pages}{535--539}.
\newblock


\bibitem[\protect\citeauthoryear{Nethercote and Seward}{Nethercote and
  Seward}{2007}]%
        {nethercote2007valgrind}
\bibfield{author}{\bibinfo{person}{Nicholas Nethercote} {and}
  \bibinfo{person}{Julian Seward}.} \bibinfo{year}{2007}\natexlab{}.
\newblock \showarticletitle{Valgrind: a framework for heavyweight dynamic
  binary instrumentation}.
\newblock \bibinfo{journal}{\emph{ACM Sigplan notices}} \bibinfo{volume}{42},
  \bibinfo{number}{6} (\bibinfo{year}{2007}), \bibinfo{pages}{89--100}.
\newblock


\bibitem[\protect\citeauthoryear{{Peng}, {Shoshitaishvili}, and {Payer}}{{Peng}
  et~al\mbox{.}}{2018}]%
        {TFUZZ}
\bibfield{author}{\bibinfo{person}{H. {Peng}}, \bibinfo{person}{Y.
  {Shoshitaishvili}}, {and} \bibinfo{person}{M. {Payer}}.}
  \bibinfo{year}{2018}\natexlab{}.
\newblock \showarticletitle{T-Fuzz: Fuzzing by Program Transformation}. In
  \bibinfo{booktitle}{\emph{2018 IEEE Symposium on Security and Privacy (SP)}}.
  \bibinfo{pages}{697--710}.
\newblock


\bibitem[\protect\citeauthoryear{Reed~Hastings}{Reed~Hastings}{1991}]%
        {reed1991purify}
\bibfield{author}{\bibinfo{person}{Bob~Joyce Reed~Hastings}.}
  \bibinfo{year}{1991}\natexlab{}.
\newblock \showarticletitle{Purify: Fast detection of memory leaks and access
  errors}. In \bibinfo{booktitle}{\emph{In Proc. of the Winter 1992 USENIX
  Conference}}. Citeseer.
\newblock


\bibitem[\protect\citeauthoryear{Ruland, Lochau, and Jakobs}{Ruland
  et~al\mbox{.}}{2020}]%
        {ruland2020hybridtiger}
\bibfield{author}{\bibinfo{person}{Sebastian Ruland}, \bibinfo{person}{Malte
  Lochau}, {and} \bibinfo{person}{Marie-Christine Jakobs}.}
  \bibinfo{year}{2020}\natexlab{}.
\newblock \showarticletitle{HybridTiger: Hybrid model checking and
  domination-based partitioning for efficient multi-goal test-suite generation
  (competition contribution)}. In \bibinfo{booktitle}{\emph{International
  Conference on Fundamental Approaches to Software Engineering}}. Springer,
  \bibinfo{pages}{520--524}.
\newblock


\bibitem[\protect\citeauthoryear{Ryabinin}{Ryabinin}{2014}]%
        {ryabinin2014ubsan}
\bibfield{author}{\bibinfo{person}{Andrey Ryabinin}.}
  \bibinfo{year}{2014}\natexlab{}.
\newblock \showarticletitle{UBSan: run-time undefined behavior sanity checker}.
\newblock \bibinfo{journal}{\emph{E-mail publi{\'e} sur la liste de
  d{\'e}veloppement du noyau Linux}}  \bibinfo{volume}{20}
  (\bibinfo{year}{2014}).
\newblock


\bibitem[\protect\citeauthoryear{Serebryany}{Serebryany}{2015}]%
        {serebryany2015libfuzzer}
\bibfield{author}{\bibinfo{person}{Kostya Serebryany}.}
  \bibinfo{year}{2015}\natexlab{}.
\newblock \showarticletitle{libFuzzer--a library for coverage-guided fuzz
  testing}.
\newblock \bibinfo{journal}{\emph{LLVM project}} (\bibinfo{year}{2015}).
\newblock


\bibitem[\protect\citeauthoryear{Serebryany, Bruening, Potapenko, and
  Vyukov}{Serebryany et~al\mbox{.}}{2012}]%
        {AddressSanitizer}
\bibfield{author}{\bibinfo{person}{Konstantin Serebryany},
  \bibinfo{person}{Derek Bruening}, \bibinfo{person}{Alexander Potapenko},
  {and} \bibinfo{person}{Dmitriy Vyukov}.} \bibinfo{year}{2012}\natexlab{}.
\newblock \showarticletitle{AddressSanitizer: A fast address sanity checker}.
  In \bibinfo{booktitle}{\emph{Presented as part of the 2012 $\{$USENIX$\}$
  Annual Technical Conference ($\{$USENIX$\}$$\{$ATC$\}$ 12)}}.
  \bibinfo{pages}{309--318}.
\newblock


\bibitem[\protect\citeauthoryear{Shoshitaishvili, Wang, Salls, Stephens,
  Polino, Dutcher, Grosen, Feng, Hauser, Kruegel,
  et~al\mbox{.}}{Shoshitaishvili et~al\mbox{.}}{2016}]%
        {angr}
\bibfield{author}{\bibinfo{person}{Yan Shoshitaishvili}, \bibinfo{person}{Ruoyu
  Wang}, \bibinfo{person}{Christopher Salls}, \bibinfo{person}{Nick Stephens},
  \bibinfo{person}{Mario Polino}, \bibinfo{person}{Andrew Dutcher},
  \bibinfo{person}{John Grosen}, \bibinfo{person}{Siji Feng},
  \bibinfo{person}{Christophe Hauser}, \bibinfo{person}{Christopher Kruegel},
  {et~al\mbox{.}}} \bibinfo{year}{2016}\natexlab{}.
\newblock \showarticletitle{Sok: (State of) The Art of War: Offensive
  techniques in binary analysis}. In \bibinfo{booktitle}{\emph{2016 IEEE
  Symposium on Security and Privacy (SP)}}. IEEE, \bibinfo{pages}{138--157}.
\newblock


\bibitem[\protect\citeauthoryear{Silver, Huang, Maddison, Guez, Sifre, Van
  Den~Driessche, Schrittwieser, Antonoglou, Panneershelvam, Lanctot,
  et~al\mbox{.}}{Silver et~al\mbox{.}}{2016}]%
        {silver2016alphago}
\bibfield{author}{\bibinfo{person}{David Silver}, \bibinfo{person}{Aja Huang},
  \bibinfo{person}{Chris~J Maddison}, \bibinfo{person}{Arthur Guez},
  \bibinfo{person}{Laurent Sifre}, \bibinfo{person}{George Van Den~Driessche},
  \bibinfo{person}{Julian Schrittwieser}, \bibinfo{person}{Ioannis Antonoglou},
  \bibinfo{person}{Veda Panneershelvam}, \bibinfo{person}{Marc Lanctot},
  {et~al\mbox{.}}} \bibinfo{year}{2016}\natexlab{}.
\newblock \showarticletitle{Mastering the game of Go with deep neural networks
  and tree search}.
\newblock \bibinfo{journal}{\emph{Nature}} \bibinfo{volume}{529},
  \bibinfo{number}{7587} (\bibinfo{year}{2016}), \bibinfo{pages}{484}.
\newblock


\bibitem[\protect\citeauthoryear{Stephens, Grosen, Salls, Dutcher, Wang,
  Corbetta, Shoshitaishvili, Kruegel, and Vigna}{Stephens
  et~al\mbox{.}}{2016}]%
        {Driller}
\bibfield{author}{\bibinfo{person}{Nick Stephens}, \bibinfo{person}{John
  Grosen}, \bibinfo{person}{Christopher Salls}, \bibinfo{person}{Andrew
  Dutcher}, \bibinfo{person}{Ruoyu Wang}, \bibinfo{person}{Jacopo Corbetta},
  \bibinfo{person}{Yan Shoshitaishvili}, \bibinfo{person}{Christopher Kruegel},
  {and} \bibinfo{person}{Giovanni Vigna}.} \bibinfo{year}{2016}\natexlab{}.
\newblock \showarticletitle{Driller: Augmenting Fuzzing Through Selective
  Symbolic Execution.}. In \bibinfo{booktitle}{\emph{NDSS}},
  Vol.~\bibinfo{volume}{16}. \bibinfo{pages}{1--16}.
\newblock


\bibitem[\protect\citeauthoryear{Szita, Chaslot, and Spronck}{Szita
  et~al\mbox{.}}{2010}]%
        {Catan}
\bibfield{author}{\bibinfo{person}{Istv{\'a}n Szita},
  \bibinfo{person}{Guillaume Chaslot}, {and} \bibinfo{person}{Pieter Spronck}.}
  \bibinfo{year}{2010}\natexlab{}.
\newblock \showarticletitle{Monte-Carlo Tree Search in Settlers of Catan}. In
  \bibinfo{booktitle}{\emph{Advances in Computer Games}},
  \bibfield{editor}{\bibinfo{person}{H.~Jaap van~den Herik} {and}
  \bibinfo{person}{Pieter Spronck}} (Eds.). \bibinfo{publisher}{Springer Berlin
  Heidelberg}, \bibinfo{address}{Berlin, Heidelberg}, \bibinfo{pages}{21--32}.
\newblock
\showISBNx{978-3-642-12993-3}


\bibitem[\protect\citeauthoryear{Van~den Broeck, Driessens, and Ramon}{Van~den
  Broeck et~al\mbox{.}}{2009}]%
        {van2009monte}
\bibfield{author}{\bibinfo{person}{Guy Van~den Broeck}, \bibinfo{person}{Kurt
  Driessens}, {and} \bibinfo{person}{Jan Ramon}.}
  \bibinfo{year}{2009}\natexlab{}.
\newblock \showarticletitle{{M}onte-{C}arlo tree search in poker using expected
  reward distributions}. In \bibinfo{booktitle}{\emph{Asian Conference on
  Machine Learning}}. Springer, \bibinfo{pages}{367--381}.
\newblock


\bibitem[\protect\citeauthoryear{Wang, Sun, Chen, Zhang, Wang, and Lin}{Wang
  et~al\mbox{.}}{2018}]%
        {MDPC}
\bibfield{author}{\bibinfo{person}{Xinyu Wang}, \bibinfo{person}{Jun Sun},
  \bibinfo{person}{Zhenbang Chen}, \bibinfo{person}{Peixin Zhang},
  \bibinfo{person}{Jingyi Wang}, {and} \bibinfo{person}{Yun Lin}.}
  \bibinfo{year}{2018}\natexlab{}.
\newblock \showarticletitle{Towards Optimal Concolic Testing}. In
  \bibinfo{booktitle}{\emph{Proceedings of the 40th International Conference on
  Software Engineering}} (Gothenburg, Sweden) \emph{(\bibinfo{series}{ICSE
  '18})}. \bibinfo{publisher}{Association for Computing Machinery},
  \bibinfo{address}{New York, NY, USA}, \bibinfo{pages}{291–302}.
\newblock
\showISBNx{9781450356381}
\urldef\tempurl%
\url{https://doi.org/10.1145/3180155.3180177}
\showDOI{\tempurl}


\bibitem[\protect\citeauthoryear{Ward and Cowling}{Ward and Cowling}{2009}]%
        {MagicTheGathering}
\bibfield{author}{\bibinfo{person}{C.D. Ward} {and} \bibinfo{person}{Peter
  Cowling}.} \bibinfo{year}{2009}\natexlab{}.
\newblock \showarticletitle{Monte Carlo search applied to card selection in
  Magic: The Gathering}. \bibinfo{pages}{9 -- 16}.
\newblock
\urldef\tempurl%
\url{https://doi.org/10.1109/CIG.2009.5286501}
\showDOI{\tempurl}


\bibitem[\protect\citeauthoryear{Xie, Ma, Juefei-Xu, Xue, Chen, Liu, Zhao, Li,
  Yin, and See}{Xie et~al\mbox{.}}{2019}]%
        {DeepHunter}
\bibfield{author}{\bibinfo{person}{Xiaofei Xie}, \bibinfo{person}{Lei Ma},
  \bibinfo{person}{Felix Juefei-Xu}, \bibinfo{person}{Minhui Xue},
  \bibinfo{person}{Hongxu Chen}, \bibinfo{person}{Yang Liu},
  \bibinfo{person}{Jianjun Zhao}, \bibinfo{person}{Bo Li},
  \bibinfo{person}{Jianxiong Yin}, {and} \bibinfo{person}{Simon See}.}
  \bibinfo{year}{2019}\natexlab{}.
\newblock \showarticletitle{DeepHunter: A Coverage-Guided Fuzz Testing
  Framework for Deep Neural Networks}. In \bibinfo{booktitle}{\emph{Proceedings
  of the 28th ACM SIGSOFT International Symposium on Software Testing and
  Analysis}} (Beijing, China) \emph{(\bibinfo{series}{ISSTA 2019})}.
  \bibinfo{publisher}{Association for Computing Machinery},
  \bibinfo{address}{New York, NY, USA}, \bibinfo{pages}{146–157}.
\newblock
\showISBNx{9781450362245}
\urldef\tempurl%
\url{https://doi.org/10.1145/3293882.3330579}
\showDOI{\tempurl}


\bibitem[\protect\citeauthoryear{{Yeh}, {Lu}, {Yeh}, and {Huang}}{{Yeh}
  et~al\mbox{.}}{2017}]%
        {MCTSSE}
\bibfield{author}{\bibinfo{person}{C. {Yeh}}, \bibinfo{person}{H. {Lu}},
  \bibinfo{person}{J. {Yeh}}, {and} \bibinfo{person}{S. {Huang}}.}
  \bibinfo{year}{2017}\natexlab{}.
\newblock \showarticletitle{Path Exploration Based on Monte Carlo Tree Search
  for Symbolic Execution}. In \bibinfo{booktitle}{\emph{2017 Conference on
  Technologies and Applications of Artificial Intelligence (TAAI)}}.
  \bibinfo{pages}{33--37}.
\newblock
\showISSN{2376-6824}
\urldef\tempurl%
\url{https://doi.org/10.1109/TAAI.2017.26}
\showDOI{\tempurl}


\bibitem[\protect\citeauthoryear{Yun, Lee, Xu, Jang, and Kim}{Yun
  et~al\mbox{.}}{2018}]%
        {yun2018qsym}
\bibfield{author}{\bibinfo{person}{Insu Yun}, \bibinfo{person}{Sangho Lee},
  \bibinfo{person}{Meng Xu}, \bibinfo{person}{Yeongjin Jang}, {and}
  \bibinfo{person}{Taesoo Kim}.} \bibinfo{year}{2018}\natexlab{}.
\newblock \showarticletitle{$\{$QSYM$\}$: A practical concolic execution engine
  tailored for hybrid fuzzing}. In \bibinfo{booktitle}{\emph{27th
  $\{$USENIX$\}$ Security Symposium ($\{$USENIX$\}$ Security 18)}}.
  \bibinfo{pages}{745--761}.
\newblock


\bibitem[\protect\citeauthoryear{Zalewski}{Zalewski}{2017}]%
        {AFL}
\bibfield{author}{\bibinfo{person}{Michal Zalewski}.}
  \bibinfo{year}{2017}\natexlab{}.
\newblock \bibinfo{title}{American fuzzy lop ({AFL}) fuzzer}.
\newblock
\newblock


\bibitem[\protect\citeauthoryear{Zhao, Duan, Yin, and Xuan}{Zhao
  et~al\mbox{.}}{2019}]%
        {DigFuzz}
\bibfield{author}{\bibinfo{person}{Lei Zhao}, \bibinfo{person}{Yue Duan},
  \bibinfo{person}{Heng Yin}, {and} \bibinfo{person}{Jifeng Xuan}.}
  \bibinfo{year}{2019}\natexlab{}.
\newblock \showarticletitle{Send Hardest Problems My Way: Probabilistic Path
  Prioritization for Hybrid Fuzzing.}. In \bibinfo{booktitle}{\emph{NDSS}}.
\newblock


\end{thebibliography}
\end{document}